\documentclass[aps, pre, reprint, floatfix, nolongbibliography, superscriptaddress, nofootinbib]{revtex4-1} 
\usepackage[utf8]{inputenc} 
\usepackage{amsfonts} 
\usepackage{amsmath} 
\usepackage{amssymb} 
\usepackage{bm} 
\usepackage{color} 
\usepackage{dcolumn} 
\usepackage{float} 
\usepackage{graphicx} 
\usepackage{mathrsfs} 
\usepackage{mathtools} 
\usepackage{scrextend} 
\usepackage[caption = false]{subfig} 
\numberwithin{equation}{section} 
\renewcommand{\theequation}{\arabic{section}.\arabic{equation}} 
\newcommand{\Eqref }[1]{Eq.~\eqref{#1}} 
\newcommand{\Eqsref}[1]{Eqs.~\eqref{#1}} 
\newcommand{\Eqnref}[1]{Equation~\eqref{#1}} 
\newcommand{\Eqnsref}[1]{Equations~\eqref{#1}} 
\newcommand{\Figref}[1]{Fig.~\ref{#1}} 
\newcommand{\Figureref}[1]{Figure~\ref{#1}} 
\newcommand{\Refref}[1]{Ref.~\onlinecite{#1}} 
\newcommand{\Referenceref}[1]{Reference~\onlinecite{#1}} 
\newcommand{\Appref}[1]{\textbf{Appendix~\ref{#1}}} 
\newcommand{\abs}[1]{ \left|{#1}\right| } % the absolute value of a number 
\newcommand{\VEC}[1]{ \left( #1 \right) } % the component form of a vector 
\newcommand{\unitvec}[1]{ \hat{ \mathbf{#1} } } % a unit vector 
\newcommand{\norm}[1]{\left\lVert{#1}\right\rVert} % the norm of a vector 
\newcommand{\p}{ \eta_{\mathrm{s}, y} } % the nondimensionalized pressure 
\newcommand{\cp}{ \eta_\text{c} } % the nondimensionalized critical pressure 
\newcommand{\CP}{ p_\text{c} } % the dimensionful critical pressure 
\newcommand{\imunit}{ \mathrm{i} } % √-1 

\DeclareMathOperator{\Laplacian}{\Delta} 
\DeclareMathOperator{\opL}{ \hat{ \mathrm{L} } } 
\DeclareMathOperator{\Vlasov}   { \Laplacian_\text{V} } % the Vlasov operator 
\DeclareMathOperator{\sinc}{ \mathrm{sinc} } 
\newcommand         {\Bessel}[1]{ \Laplacian_{#1}     } % the Bessel operator 
\DeclareMathOperator{\diag}{diag} 
\DeclareMathOperator{\tr}  {tr} 
\newcommand{\diagMatrix}[1]{ \diag\left\{#1\right\} } 
\newcommand{\df}[2][]{ \mathrm {d}^{#1}{#2} } % a differential (form) 
\newcommand{\odfrac}[3][]{ \frac{ \mathrm{d}^{#1}{#2} }{ { \mathrm{d}{#3} }^{#1} } } % the ordinary derivative of a function 
\newcommand{\pdfrac}[3][]{ \frac{ \partial  ^{#1}{#2} }{ { \partial  {#3} }^{#1} } } % the partial  derivative of a function 
\newcommand{\mpdfrac}[3] { \frac{ \partial  ^2   {#1} }{ \partial{#2}\, \partial{#3} } } 
\DeclareMathOperator{\shape}{ \mathcal{K} } 
 
\newcommand{\textAnd}{\quad \text{and} \quad} 
 
\newcommand{\EllipticK}[1]{ \mathrm{K}\left(#1\right) } 
\newcommand{\EllipticF}[2]{ \mathrm{F}\left(#1\left|#2\right.\right) } 
\newcommand{\LH} {L'H\^{o}pital} 
\newcommand{\FvK}{F\"{o}ppl-von K\'{a}rm\'{a}n} 
\newcommand{\Reqeq}{ \stackrel{!}{=} } 
\def\COMSOL{\texttt{COMSOL}}
\def\eff  { \mathrm{eff} } 
\def\Huber{\mathrm H}

\begin{document} 
\title{A Geometric Mapping from Rectilinear Material Orthotropy to Isotropy: Insights to Plates and Shells} 
\author{Wenqian Sun} 
\email[]{wenqians@uoregon.edu} 
\affiliation{Institute for Fundamental Science and Department of Physics, University of Oregon, Eugene, Oregon 97403, USA} 
\author{Cody Rasmussen} 
\affiliation{Institute for Fundamental Science and Department of Physics, University of Oregon, Eugene, Oregon 97403, USA} 
\author{Roman Vetter} 
\affiliation{Computational Physics for Engineering Materials, ETH Zurich, 8093 Zurich, Switzerland}
\affiliation{Current address: Department of Biosystems Science and Engineering, ETH Zurich, 4058 Basel, Switzerland} 
\author{Jayson Paulose} 
\email[]{jpaulose@uoregon.edu} 
\affiliation{Institute for Fundamental Science and Department of Physics, University of Oregon, Eugene, Oregon 97403, USA} 
\affiliation{Material Science Institute,                                  University of Oregon, Eugene, Oregon 97403, USA} 
\date{\today} 
\begin{abstract}

Orthotropic shell structures are ubiquitous in biology and engineering, from bacterial cell walls to reinforced domes.
We present a rescaling transformation that maps an orthotropic shallow shell to an isotropic one with a different local geometry.
The mapping is applicable to any shell section for which the material orthotropy directions match the principal curvature directions, assuming a commonly used form for the orthotropic shear modulus.
Using the rescaling transformation, we derive exact expressions for the buckling pressure as well as the linear indentation response of orthotropic cylinders and general ellipsoids of revolution, which we verify against numerical simulations.
Our analysis disentangles the separate contributions of geometric and material anisotropy to shell rigidity.
In particular, we identify the geometric mean of orthotropic elastic constants as the key quantifier of material stiffness, playing a role akin to the Gaussian curvature which captures the geometric stiffness contribution. 
Besides providing insights into the mechanical response of orthotropic shells, our work rigorously establishes the validity of isotropic approximations to orthotropic shells and also identifies situations in which these approximations might fail.

\end{abstract} \maketitle 
\section{Introduction} \label{sec: introduction} 
Isotropic elasticity, which assumes material properties that are independent of direction, provides a tractable and convenient description of many everyday mechanical phenomena.
However, direction-dependent mechanical properties are the rule rather than the exception in natural materials, from muscle tissue~\cite{Xu2022_Muscle} and wood~\cite{Bucur1988_Wood} to the cell walls of bacteria~\cite{Thwaites1991_Bacteria} and plants~\cite{Baskin2005}. 
The mechanical anisotropy is typically a result of high-strength filaments or fibers within these materials that are oriented in a particular direction, strengthening the direction and hence breaking the material rotational symmetry (i.e., isotropy)~\cite{Timoshenko_Introductory}. 
In the technological realm,  composite materials with directional reinforcements such as plywood~\cite{Anisotropy_Text} and corrugated materials~\cite{Orange_Bible} are used to build structures that are mechanically strong and resilient in desired directions; the elastic description of these structures at length scales larger than the reinforcement features also requires anisotropic material parameters.

Thin-walled elastic structures, or shells, provide a rich setting for interesting elastic phenomena that arise from the interplay of material anisotropy and geometry.
For example, a thin cylindrical shell whose inner wall is wrapped helically by polymer fibers can develop into a spiral shape upon expansion, which has been proposed as a model for bacterial growth~\cite{Wolgemuth2005}. 
In engineered shell structures, closely spaced ribs provide strength in high-stress directions with minimal addition of material in e.g., masonry domes~\cite{Bayraktar2023} and pressure vessels~\cite{Buchert1965_Buckling}; these directional reinforcements strongly influence the failure modes of the shells~\cite{Herrmann2019} and can generate multistability in shell conformations~\cite{Guest2006a,Seffen2007,Vidoli2008_Tristability,Sobota2019_Ortho_Shapes}.
Besides its fundamental interest to mechanics, the interplay of anisotropic elasticity, shell geometry and external loading is crucial to our understanding of cell biophysics as well as to structural engineering. 

\newsavebox{\myimage} 
\begin{figure*}[tb] 
\includegraphics{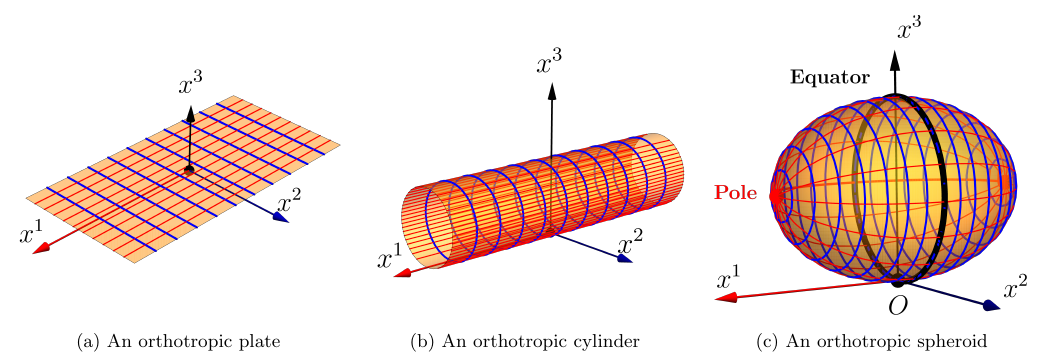}
\caption{
Plates and shells with local rectilinear orthotropy. 
The two material orthotropic directions are marked by different colors, the $x^1$-direction by red and $x^2$-direction by blue. 
For all the three structures, these two directions are also the principal directions of curvature. 
In this paper, we only consider rectilinear orthotropy---shell sections that locally look like (a); shell regions that are curvilinearly orthotropic, e.g., the poles of the orthotropic spheroid, are beyond the scope of this study. 
For curved shells, (b) and (c), we take the $x^2$-direction to be the azimuthal direction, so $R_2$ denotes the equatorial radius of the spheroid.
} 
\label{fig: orthotropic structures} 
\end{figure*} 

One obstacle to building a fundamental understanding of shells with anisotropic elasticity is that the reduction in material symmetries makes the governing differential equations more challenging to solve. 
For instance, twenty-one independent elastic constants are needed to fully characterize a three-dimensional anisotropic material (while only two are needed in the isotropic case)~\cite{LL_Elasticity,Orange_Bible}. 
Here, we study a particular type of material anisotropy---two-dimensional orthotropic materials (or equivalently thin three-dimensional transversely isotropic materials). 
Such materials have different elastic properties along two orthogonal in-plane directions, one of which has the same material composition as the material thickness direction~\cite{Anisotropy_Text}, see \Figref{fig: orthotropic structures}.
This form of anisotropy provides a good approximation to engineered thin-walled structures such as fiber-reinforced shells~\cite{Orange_Bible, Becker1958_Stiffened_cylinders} and shells with linear corrugations~\cite{Xia2012}. 
Orthotropic elasticity also arises as a natural consequence of the growth mechanism of rod-shaped bacterial cell walls, in which stiff carbohydrate chains are laid down by molecular complexes along the circumferential direction~\cite{Teeffelen2011,Garner2011,Dominguez-Escobar2011} breaking local material symmetry~\cite{Wang2012a,Zhang2021}. 
Orthotropy therefore serves as a tractable yet relevant model for assessing the influence of material anisotropy on shell mechanics.
Nevertheless, the lowered symmetry of the governing shell equations has typically favored numerical analyses of orthotropic shell response~\cite{Penzes1969,Penzes1973,Indian_Ortho_Sph_Cap,Uddin1987,Semenov2016,Herrmann2019}, although a few analytical results exist for buckling thresholds~\cite{Becker1968_Ship_Research} and multistability criteria~\cite{Guest2006a,Seffen2007,Vidoli2008_Tristability,Sobota2019_Ortho_Shapes} of orthotropic shells.

In this work, we establish an exact mapping between orthotropic and isotropic shells, and apply this mapping to generate analytical results for the local mechanical response of orthotropic shells.
Specifically,  we will demonstrate that although the orthotropic materials still have a reduced symmetry compared to isotropic materials, they become effectively isotropic under an appropriately chosen coordinate transformation. 
A specific version of this isotropy-orthotropy equivalence have been recognized for linear orthotropic plate equations~\cite{Timoshenko_Bible, Panc1975}; here, we rigorously establish the equivalence using the tensor formulation of elasticity, and generalize it to nonlinear deflections of curved shell sections described by shallow-shell theory~\cite{Orange_Bible}.
Under the aforementioned coordinate transformation, the orthotropic shallow-shell equations are mapped to a system of equations describing a shallow shell made of an isotropic material, but with different geometric parameters.
We apply the transformation to study local mechanical properties---linear response to an indentation force~\cite{Vaziri2008,VellaSph,Vella2012Spheroid,Lazarus2012a} and buckling load---of thin-walled structures that are made of orthotropic materials. 
These local mechanical properties have recently been established rigorously for isotropic shells with arbitrary curvatures and pressures~\cite{VellaSph, Vella2012Spheroid,Wenqian2021}; however, to our knowledge, our mapping enables the first analytical results for the local response of orthotropic shells. \section{Background} \label{sec: background} 
We start with the elastic description for a two-dimensional\footnote{
Realistically, every material has a finite thickness and is hence three-dimensional. 
The materials considered here are effectively two-dimensional, i.e., so thin that the Kirchhoff-Love hypothesis~\cite{Orange_Bible} applies.} orthotropic material, which relates local strains to local stresses via a stiffness tensor. 
Let $u_{\alpha\beta}$ be the covariant components of the strain tensor, and let $\sigma^{\alpha\beta}$ denote the contravariant components of the stress tensor; $( \alpha, \beta \in \{1, 2\} )$. 
The generalized Hooke's law for an orthotropic material is: 
$u_{\alpha\beta} = C_{\alpha\beta\gamma\delta}\, \sigma^{\gamma\delta}$, 
where $\mathbf{C}$ is the rank-four stiffness tensor~\cite{LL_Elasticity}. (The Einstein convention of summation over repeated upper and lower indices is implied throughout the paper.) 
In Voigt notation, this reads~\cite{Anisotropy_Text} 
\begin{equation} 
\begin{pmatrix} 
u_{11} \\[0.25 em] 
u_{22} \\[0.25 em] 
u_{12} 
\end{pmatrix} 
= 
\begin{pmatrix} 
 \cfrac{1}              {E_1} & -\cfrac{ \upsilon_{21} }{E_2} & 0 \\[1 em] 
-\cfrac{ \upsilon_{12} }{E_1} &  \cfrac{1}              {E_2} & 0 \\[1 em] 
0                             & 0                             & \cfrac{1}{ 2G_{12} } 
\end{pmatrix} 
\begin{pmatrix} 
\sigma^{11} \\[0.25 em] 
\sigma^{22} \\[0.25 em] 
\sigma^{12} 
\end{pmatrix} 
, 
\label{eqn: the generalized Hooke's law} 
\end{equation} 
where $E_\alpha$ and $\upsilon_{\alpha\beta}$ $(\alpha \not= \beta)$ denote Young's moduli and Poisson's ratios along the two orthogonal directions, respectively. 
In this paper, we consider the common case where these elastic constants are all positive. 
By Betti's reciprocal theorem~\cite{Orange_Bible}, 
\begin{equation} 
\frac{ \upsilon_{21} }{E_2} = \frac{ \upsilon_{12} }{E_1}. 
\label{eqn: the symmetry of the stiffness tensor} 
\end{equation} 
We can accordingly define a parameter which characterizes the degree of material anisotropy: 
\begin{equation} 
\lambda \coloneqq \frac{E_1}{E_2} = \frac{ \upsilon_{12} }{ \upsilon_{21} } > 0. 
\label{eqn: the anisotropy parameter} 
\end{equation} 
The positive definiteness of the stiffness matrix in Voigt notation, $\det\left(C_{ [\alpha\beta][\gamma\delta] }\right) > 0$, imposes an upper bound for the anisotropy parameter: $\lambda < \frac{1}{\upsilon_{12}^2}$. 
The Poisson's ratio $\upsilon_{12}$ can in principle be zero~\cite{Lempriere1968_Poisson_ratio}; as a result, $\lambda \in (0, \infty)$ (recall that we assume $\upsilon_{12}, \upsilon_{21} > 0$). 
The inverse of $\lambda$, $\frac{1}{\lambda} \coloneqq \frac{E_2}{E_1}$, also has the same range of values. 
In practice, given a general two-dimensional orthotropic material, one is free to call the first direction either of the two principal directions of the stiffness tensor $\mathbf C$ and hence use either $\lambda$ or $\frac{1}{\lambda}$ to characterize the degree of material anisotropy. 
\par 
Because of \Eqref{eqn: the symmetry of the stiffness tensor}, one only needs four independent parameters to fully characterize a two-dimensional orthotropic material. 
We choose the four to be 
$       E_\eff \coloneqq \sqrt{ E_1          E_2           }$, 
$\upsilon_\eff \coloneqq \sqrt{ \upsilon_{12}\upsilon_{21} }$, 
$\lambda$ and $G_{12}$. (We will see the reason for this choice in \Eqref{eqn: the Huber form} and \textbf{\ref{subsec: a rescaling transformation}}.) 
The elastic constant $G_{12}$ is the material's in-plane shear modulus and is, in general, an independent quantity. 
However, in practice it is closely related to the Young's moduli in the orthotropic directions. 
To eliminate this degree of freedom, M. T. Huber proposed the following form for $G_{12}$~\cite{Huber1923}, 
\begin{equation} 
G_{12} \Reqeq    G_\Huber 
       \coloneqq \frac{ E_\eff        }{ 2( 1 + \upsilon_\eff                       ) } 
               = \frac{ \sqrt{E_1E_2} }{ 2( 1 + \sqrt{ \upsilon_{12}\upsilon_{21} } ) }, 
\label{eqn: the Huber form} 
\end{equation} 
substituting the \emph{geometric means} of the anisotropic elastic constants as effective constants into the expression of the shear modulus of an isotropic material. 
The Huber form for the orthotropic shear modulus has been accepted and widely employed in both analytical and numerical calculations~\cite{Orange_Bible, Timoshenko_Bible, Herrmann2019, Cheng1984Composite_Cylinders, Paschero2009Ortho_Cylinders}. 
Panc demonstrated, based on theoretical arguments, that for orthotropic materials, the Huber form may be used as an approximation~\cite{Panc1975}. 
Cheng and He further argued that although the Huber form is itself inaccurate for fiber-reinforced composite materials, it can still yield accurate analytical results when substituted in governing differential equations of shell theory (at least for cylinders)~\cite{Cheng1984Composite_Cylinders}. 
\par 
The following result section is structured as follows. 
In \textbf{\ref{subsec: a rescaling transformation}}, we introduce the main result of this paper--the rescaling transformation which shows that an orthotropic two-dimensional material becomes effectively isotropic if we use a rescaled Cartesian coordinate system. 
In \textbf{\ref{subsec: shallow-shell equations}}, we exploit the use of the transformation in shallow-shell systems. 
We demonstrate that the general Donnell-Mushtari-Vlasov (DMV) equations, the governing equations in the shallow-shell theory, are covariant under the transformation and use the transformation to derive the DMV equations for orthotropic shells in a physically transparent manner. 
In \textbf{\ref{subsec: the local indentation stiffness}} and \textbf{\ref{subsec: the buckling load}}, by solving these equations, we obtain the indentation stiffness and buckling pressure of orthotropic ellipsoids and cylinders. \section{Results} \label{sec: results} 
\subsection{A Rescaling Transformation} \label{subsec: a rescaling transformation} 
\paragraph{Transformation Step 1.} \label{par: step 1 of the rescaling transformation} 
We first notice that with the Huber form (\Eqref{eqn: the Huber form}), \Eqref{eqn: the generalized Hooke's law} can be rewritten, in terms of the effective elastic constants and the anisotropy parameter $\lambda$, as 
\begin{equation} 
\begin{pmatrix} 
           \sqrt[4]{\lambda}  u_{11} \\[0.25 em] 
\cfrac{1}{ \sqrt[4]{\lambda} }u_{22} \\[1 em] 
                              u_{12} 
\end{pmatrix} 
= 
\begin{pmatrix} 
 \cfrac{1}            {E_\eff} & -\cfrac{\upsilon_\eff}{E_\eff} & 0 \\[1 em] 
-\cfrac{\upsilon_\eff}{E_\eff} &  \cfrac{1}            {E_\eff} & 0 \\[1 em] 
 0                             &  0                             & \cfrac{1 + \upsilon_\eff}{E_\eff} 
\end{pmatrix} 
\begin{pmatrix} 
\cfrac{1}{ \sqrt[4]{\lambda} }\sigma^{11} \\[1 em] 
           \sqrt[4]{\lambda}  \sigma^{22} \\[0.25 em] 
                              \sigma^{12} 
\end{pmatrix} 
. 
\label{eqn: transformation step 1} 
\end{equation} 
The stiffness matrix now takes the form of that for an isotropic material with elastic constants $\{E_\eff, \upsilon_\eff\}$~\cite{Anisotropy_Text}. 
\Eqnref{eqn: transformation step 1} in fact implies that an orthotropic material can be treated as isotropic if we rescale physical quantities in a systematic way. 
This can be seen more clearly using tensors. 
In tensor notation, \Eqref{eqn: transformation step 1} can be written as 
$
  u     _{\alpha'\beta'              } 
= C     _{\alpha'\beta'\gamma'\delta'}\, 
  \sigma^{             \gamma'\delta'} 
$. 
Primed indices are used here to denote the transformed tensor components: 
\refstepcounter{equation} \label{eqn: rescaled tensor components} 
\begin{equation} 
  u_{\alpha'\beta'} 
= {\Lambda^\alpha}_{\alpha'}\, 
  {\Lambda^\beta }_{\beta' }\, 
  u_{\alpha\beta}, 
\tag{\theequation, a} 
\label{eqn: rescaled strain tensor components} 
\end{equation} 

\begin{equation} 
  \sigma^{\alpha'\beta'} 
= { \Lambda^{\alpha'} }_\alpha\, 
  { \Lambda^{\beta' } }_\beta \, 
  \sigma^{\alpha\beta} 
\tag{\theequation, b} 
\label{eqn: rescaled stress tensor components} 
\end{equation} 
and 
\begin{equation} 
  C_{\alpha'\beta'\gamma'\delta'} 
= {\Lambda^\alpha}_{\alpha'}\, 
  {\Lambda^\beta }_{\beta' }\, 
  {\Lambda^\gamma}_{\gamma'}\, 
  {\Lambda^\delta}_{\delta'}\, 
  C_{\alpha\beta\gamma\delta}, 
\tag{\theequation, c} 
\label{eqn: rescaled stiffness tensor components} 
\end{equation} 
where $\left( {\Lambda^i}_{i'} \right) \coloneqq \diagMatrix{\sqrt[8]{\lambda}, \frac{1}{ \sqrt[8]{\lambda} }, 1}$, and 
${\Lambda^i}_{i'}\, { \Lambda^{i'} }_j = \delta^i_j$ with $\delta^i_j$ the Kronecker delta. (Latin indices run from $1$ to $3$, while Greek indices only take on values $1$ and $2$.) 
That is, when written in terms of the rescaled tensor components, the anisotropic Hooke's law takes the isotropic form. 
This shows that the orthotropic material becomes effectively isotropic if we hide the material anisotropy by rescaling the strain and the stress components. 
We note that this rescaling transformation preserves the elastic energy density: 
$
  u_{\alpha \beta }\, \sigma^{\alpha \beta } 
= u_{\alpha'\beta'}\, \sigma^{\alpha'\beta'} 
$. 
\par 
In fact, the total elastic energy is also invariant under the transformation. 
\Eqnsref{eqn: rescaled tensor components} hint at the following coordinate transformation: 
\begin{equation} 
x^{i'} = { \Lambda^{i'} }_j\, x^j. 
\label{eqn: the underlying coordinate transformation} 
\end{equation} 
Let $g_{ij}$ denote the unscaled components of the metric tensor; its rescaled components can then be computed: 
$g_{i' j'} = {\Lambda^i}_{i'}\, {\Lambda^j}_{j'}\, g_{ij}$. 
Note that $\det( g_{i' j'} ) = \det( g_{ij} )$, since $\det( {\Lambda^i}_{i'} ) = 1$. 
This further implies that 
\begin{widetext} 
\begin{equation} 
U = \frac{1}{2}\int_ \mathcal{M}   \sqrt{ \det( g_{ij}    ) }\, \df[2]{ \mathbf{x} }\, 
    u_{\alpha \beta }\, \sigma^{\alpha \beta } 
  = \frac{1}{2}\int_{\mathcal{M}'} \sqrt{ \det( g_{i' j'} ) }\, \df[2]{ \mathbf{x}'}\, 
    u_{\alpha'\beta'}\, \sigma^{\alpha'\beta'}, 
\label{eqn: invariance of total elastic energy} 
\end{equation} 
\end{widetext} 
i.e., the total energy is preserved. 
\paragraph{Transformation Step 2.} \label{par: step 2 of the rescaling transformation} 
The strain tensor is related to deformation displacement fields via the so-called strain-displacement relations. 
We are now going to demonstrate that the rescaling transformation is compatible with these relations. 
Since all materials are three-dimensional, we will use the relations for a thin curved material (i.e., a shallow shell) that satisfies the Kirchhoff-Love hypothesis~\cite{Orange_Bible}, which basically assumes that no deformation occurs along the thickness direction. 
\par 
For such a shell, the Green-Lagrange strain tensor is given, in terms of two in-plane phonon fields $u_\alpha( \mathbf{x} )$ and one out-of-plane deformation field $u_3( \mathbf{x} )$, by~\cite{Koiter_notes, Orange_Bible} 
\begin{align} 
\begin{split} 
u_{\alpha\beta} & = \frac{1}{2}\left( 
                    \partial_\alpha{u_\beta } 
                  + \partial_\beta {u_\alpha} 
                  + \partial_\alpha{u_3     } 
                  \cdot 
                    \partial_\beta {u_3     } 
                    \right) 
                  - \shape_{\alpha\beta}^0u_3 
                    - \\[0.25 em] 
                & \quad - 
                    x_3\, 
                    \partial_\alpha{ 
                    \partial_\beta {u_3} 
                    }, 
\end{split} 
\label{eqn: strain-displacement relation} 
\end{align} 
where $\partial_\alpha \equiv \pdfrac{}{x^\alpha}$, and $(\shape_{\alpha\beta}^0) = \diagMatrix{\kappa_1, \kappa_2}$ is the extrinsic curvature tensor that encodes the two local principal curvatures of the material's undeformed middle surface. 
For a sphere with radius $R$, $\shape_{\alpha\beta}^0 = \frac{1}{R}\, \delta_\beta^\alpha$, while a cylinder of the same radius has $\shape_{\alpha\beta}^0 = \frac{1}{R}\, \delta_\alpha^1\, \delta_\beta^1$ (or 
$\shape_{\alpha\beta}^0 = \frac{1}{R}\, \delta_\alpha^2\, \delta_\beta^2$). 
The last term in \Eqref{eqn: strain-displacement relation} is the bending strain~\cite{Orange_Bible}, where $x_3$ denotes the distance away from the middle surface. 
\par 
The rescaled components can then be written, using \Eqref{eqn: rescaled strain tensor components}, as 
\begin{widetext} 
\begin{align} 
\begin{split} 
u_{\alpha'\beta'} & = \frac{1}{2}\big[ 
                      \left( {\Lambda^\alpha}_{\alpha'}\, \partial_\alpha \right)\left( {\Lambda^\beta }_{\beta' }\, u_\beta  \right) 
                    + \left( {\Lambda^\beta }_{\beta' }\, \partial_\beta  \right)\left( {\Lambda^\alpha}_{\alpha'}\, u_\alpha \right) 
                    + \left( {\Lambda^\alpha}_{\alpha'}\, \partial_\alpha \right){u_3} 
                    \cdot 
                      \left( {\Lambda^\beta }_{\beta' }\, \partial_\beta  \right){u_3} 
                      \big] 
                    - \left( 
                      {\Lambda^\alpha}_{\alpha'}\, 
                      {\Lambda^\beta }_{\beta' }\, 
                      \shape_{\alpha\beta}^0 
                      \right) 
                      u_3 
                      - \\[0.25 em] 
                  & \quad - 
                      x_3\, 
                      \left( {\Lambda^\alpha}_{\alpha'}\, \partial_\alpha \right) 
                      \left( {\Lambda^\beta }_{\beta' }\, \partial_\beta  \right) 
                      {u_3} 
                      \\[0.25 em] 
                  & = \frac{1}{2}\left( 
                      \partial_{\alpha'}{ u_{\beta' } } 
                    + \partial_{\beta' }{ u_{\alpha'} } 
                    + \partial_{\alpha'}{ u_{3'     } } 
                    \cdot 
                      \partial_{\beta' }{ u_{3'     } } 
                      \right) 
                    - \shape_{\alpha'\beta'}^0u_{3'} 
                    - x_{3'}\, 
                      \partial_{\alpha'}{ 
                      \partial_{\beta' }{ u_{3'} } 
                      }. 
\label{eqn: rescaled strain-displacement relation} 
\end{split} 
\end{align} 
\end{widetext} 
For the sake of consistency, we have written in the above equation $x_{3'} = {\Lambda^i}_{3'}\, x_i = x_3$ and $u_{3'} = {\Lambda^i}_{3'}\, u_i = u_3$. 
Note that both the coordinate and the displacement along the thickness direction remain unrescaled. 
\par 
\Eqnsref{eqn: strain-displacement relation} and \eqref{eqn: rescaled strain-displacement relation} take exactly the same form. 
This means that rescaling the underlying deformation displacement fields can indeed lead to the rescaled strain-tensor field, indicating the compatibility between the rescaling transformation and the strain-displacement relations. 
The only difference between the two equations is that the extrinsic curvature tensor, in the rescaled coordinate system, now becomes 
\begin{equation} 
(\shape_{\alpha'\beta'}^0) = \diagMatrix{ 
                             \kappa_{1'}, 
                             \kappa_{2'} 
                             } 
                           \coloneqq 
                             \diagMatrix{ 
                                       \sqrt[4]{\lambda}  \kappa_1, 
                             \frac{1}{ \sqrt[4]{\lambda} }\kappa_2 
                             }. 
\label{eqn: rescaled extrinsic curvature tensor} 
\end{equation} 
This shows that the material's middle surface has a different local geometry in the rescaled coordinate system. 
For example, a sphere with radius $R$ becomes locally an ellipsoid with principle radii of curvature $\frac{1}{ \sqrt[4]{\lambda} }R$ and $\sqrt[4]{\lambda}R$. 
Nonetheless, note that the local Gaussian curvature remains unchanged: 
\begin{equation} 
K \equiv \det(\shape_{\alpha \beta }^0) 
       = \kappa_1\kappa_2 
       = \det(\shape_{\alpha'\beta'}^0) 
  \equiv K'; 
\label{eqn: invariance of Gaussian curvature} 
\end{equation} 
while the other invariant of the extrinsic curvature tensor, the local mean curvature $H \equiv \frac{1}{2}\tr(\shape_{\alpha\beta}^0)$ does not remain invariant under the rescaling: 
\begin{equation} 
       H 
     = \frac{1}{2}\left(\kappa_ 1   + \kappa_ 2  \right) 
\not = \frac{1}{2}\left(\kappa_{1'} + \kappa_{2'}\right) 
     = \frac{1}{2}\tr(\shape_{\alpha'\beta'}^0) 
  \equiv H'. 
\label{eqn: mean curvature under rescaling} 
\end{equation} 
\par 
To sum up, we have established a curious rescaling transformation (\Eqsref{eqn: the underlying coordinate transformation} and \eqref{eqn: rescaled tensor components}), assuming the Huber form for the orthotropic in-plane shear modulus. 
The transformation implies that under certain circumstances, such as cases where shear deformations are negligible, an orthotropic material can exhibit similar elastic behaviors as an isotropic one with different local geometrical properties. 
\par 
It should be pointed out that we have made a couple of assumptions when establishing the above equivalence relationship. 
The first one is that the material-orthotropy pattern must be \emph{rectilinear} (i.e., can be characterized locally by a Cartesian coordinate system), not \emph{curvilinear}, and the two orthogonal directions have to coincide with directions of local principal curvatures (see \Figref{fig: orthotropic structures}). 
Also, the form of the strain tensor, \Eqref{eqn: strain-displacement relation}, implicitly requires that the deformation displacements vary rapidly, on the scale of curvature radii, along the principal directions, i.e., 
$\abs{ \frac{1}{u_\beta}\partial_\alpha{u_\beta} } \gg \frac{1}{ \min\left\{R_1, R_2\right\} }$, 
where $R_\alpha \equiv \frac{1}{\kappa_\alpha}$~\cite{Orange_Bible}. 
Given that $\lambda$ is of order one, which implies that $\sqrt[8]{\lambda}$ is approximately unity, the same requirement in the rescaled coordinate system, $\abs{ \frac{1}{ u_{\beta'} }\partial_{\alpha'}{ u_{\beta'} } } \gg \frac{1}{ \min\left\{R_{1'}, R_{2'}\right\} }$, can accordingly still be satisfied. 
In the context of thin shells, this means that a shallow shell remains shallow after getting rescaled. 
\par 
We now move on to discuss several implications of the established equivalence relationship. 
The first and foremost perhaps is that we can effortlessly obtain, without performing any functional analysis, the equation of equilibrium and the compatibility equation for an orthotropic doubly-curved shallow shell. 
The equations will be presented in a covariant way, in tensor notation, to illustrate that they are form-invariant under the rescaling transformation. 
\subsection{Equations of the Shallow-Shell Theory} \label{subsec: shallow-shell equations} 
Recall that we have demonstrated that an orthotropic shallow shell with the set of parameters $\{E_1, \upsilon_{21}, \lambda; R_1, R_2\}$ shares the same total-energy functional with an \emph{isotropic} one whose corresponding parameters are given by 
$
\left\{ 
E_\eff        \equiv \sqrt{ E_1          E_2           }, 
\upsilon_\eff \equiv \sqrt{ \upsilon_{12}\upsilon_{21} }; 
R_{1'} \equiv \frac{R_1}{ \sqrt[4]{\lambda} }, 
R_{2'} \equiv \sqrt[4]{\lambda}R_2 
\right\} 
$. 
Since minimizing the total-energy functional gives the equation of equilibrium (EOE), we conclude that the EOE for the orthotropic shell will be the same as the corresponding isotropic EOE when written in terms of rescaled quantities: 
\refstepcounter{equation} \label{eqn: equations of SST} 
\begin{equation} 
  D'\opL'{ u_{3'} } 
+ \sigma^{\alpha'\beta'}t'\left( 
  \shape_{\alpha'\beta'}^0 
- \partial_{\alpha'}{ 
  \partial_{\beta' }{ u_{3'} } 
  } 
  \right) 
= p'\left(x^{\alpha'}\right), 
\tag{\theequation, a} 
\label{eqn: the EOE} 
\end{equation} 
where $D' \coloneqq \frac{E_\eff t^3}{ 12(1 - \upsilon_\eff^2) }$ is the effective bending modulus; $t = t'$ the shell thickness; and $p'$ describes the load applied to the shell. 
The operator $\opL'{}$ denotes the linear differential operator $\pdfrac[4]{}{x'} + 2\pdfrac[2]{}{x'}\pdfrac[2]{}{y'} + \pdfrac[4]{}{y'}$.\footnote{
The fully covariant way of writing the operator is 
$
D^{\alpha\beta\gamma\delta}\, 
\partial_\alpha{ 
\partial_\beta { 
\partial_\gamma{ 
\partial_\delta{ 
} } } } 
$, where $\mathbf{D}$ denotes the bending-stiffness tensor: In Voigt notation, 
\begin{align*} 
    \left(D^{ [\alpha\beta][\gamma\delta] }\right) 
& = \begin{pmatrix} 
    D^{1111} & D^{1122} & D^{1112} & D^{1121} \\[0.25 em] 
    D^{2211} & D^{2222} & D^{2212} & D^{2221} \\[0.25 em] 
    D^{1211} & D^{1222} & D^{1212} & D^{1221} \\[0.25 em] 
    D^{2111} & D^{2122} & D^{2112} & D^{2121} 
    \end{pmatrix} 
    \\[0.25 em] 
& = D' 
    \begin{pmatrix} 
    \sqrt{\lambda} & \upsilon_\eff              & 0                           & 0 \\[0.25 em] 
    \upsilon_\eff  & \frac{1}{ \sqrt{\lambda} } & 0                           & 0 \\[0.25 em] 
    0              & 0                          & \frac{1 - \upsilon_\eff}{2} & \frac{1 - \upsilon_\eff}{2} \\[0.25 em] 
    0              & 0                          & \frac{1 - \upsilon_\eff}{2} & \frac{1 - \upsilon_\eff}{2} \\[0.25 em] 
    \end{pmatrix}, 
\end{align*} 
again using the Huber form.} 
Note that in spite of its appearance, $\opL'{}$ is in fact \emph{not} the biharmonic operator in the rescaled coordinate system.\footnote{
The Laplacian operator, or rather the Laplace-Beltrami operator, in the rescaled coordinate system, which is non-Euclidean, is 
$
\Laplacian'{} \equiv \frac{1}{ \sqrt{g'} }\partial_{\alpha'}\left( 
                               \sqrt{g'} 
                     g^{\alpha'\beta'}\,  \partial_{\beta' }{} 
                     \right) 
                   = \frac{1}{ \sqrt[4]{\lambda} }\pdfrac[2]{}{x'} 
                   +           \sqrt[4]{\lambda}  \pdfrac[2]{}{y'} 
$, where $g' \equiv \det( g_{\alpha'\beta'} )$. 
} 
\par 
Recall that the strain-displacement relations (\Eqref{eqn: strain-displacement relation}) also take the same form in both coordinate systems. 
By the same reasoning, the fact that the compatibility equation stems from strain-displacement relations~\cite{Paulose2013SoftSpots} implies that for the orthotropic shell, the compatibility equation is given by 
\begin{equation} 
  \frac{1}{Y'}\opL'{\Phi'} 
= \varepsilon^{\alpha'\gamma'} 
  \varepsilon^{\beta' \delta'} 
  \partial_{\gamma'}{ 
  \partial_{\delta'}{ u_{3'} } 
  }\left( 
  \shape_{\alpha'\beta'}^0 
- \frac{1}{2} 
  \partial_{\alpha'}{ 
  \partial_{\beta' }{ u_{3'} } 
  } 
  \right), 
\tag{\theequation, b} 
\label{eqn: the compatibility equation} 
\end{equation} 
where $Y' \coloneqq E_\eff t$ is the effective two-dimensional Young's modulus. 
The Airy stress function $\Phi'$ is a scalar field and hence unrescaled, i.e., $\Phi'\left(x^{\alpha'}\right) = \Phi(x^\alpha)$. 
It is related to the rescaled stress components in the following way: 
\begin{equation} 
\sigma^{\alpha'\beta'}t' = \varepsilon^{\alpha'\gamma'} 
                           \varepsilon^{\beta' \delta'} 
                           \partial_{\gamma'}{ 
                           \partial_{\delta'}{\Phi'} 
                           }, 
\label{eqn: Airy stress function and stress components} 
\end{equation} 
where $\varepsilon^{\alpha'\beta'}$ is the rescaled components of the two-dimensional alternating tensor. 
\par 
\Eqnsref{eqn: equations of SST} are the nonlinear shallow-shell equations for the orthotropic shell. 
The linearized version can be obtained via the procedure outlined in \Refref{Orange_Bible}; the results are shown below: 
\refstepcounter{equation} \label{eqn: linearized equations of SST} 
\begin{align*} 
& D'\opL'{ u_{3'} } 
+ \sigma  ^{\alpha'\beta'}t' 
  \shape  _{\alpha'\beta'}^0 
- \sigma_0^{\alpha'\beta'}t' 
  \partial_{\alpha'}{ 
  \partial_{\beta' }{ u_{3'} } 
  } 
= 0 
  \tag{\theequation, a} 
  \label{eqn: the linearized EOE} 
  \\[0.25 em] 
& Y' 
  \varepsilon^{\alpha'\gamma'} 
  \varepsilon^{\beta' \delta'} 
  \shape     _{\alpha'\beta' }^0 
  \partial_{\gamma'}{ 
  \partial_{\delta'}{ u_{3'} } 
  } 
= \opL'{\Phi'}, 
  \tag{\theequation, b} 
  \label{eqn: the linearized compatibility equation} 
\end{align*} 
where $\sigma_0^{\alpha'\beta'}$ denotes the rescaled prestress components. 
\Eqnsref{eqn: linearized equations of SST} are consistent with known expressions in the literature~\cite{Nemeth1994SSEquations}. 
Equations written in terms of unrescaled quantities without tensor notation can be found in \Appref{appendix: unrescaled equations}. 
\par 
The linearized equations can be employed to study the local indentation stiffness of a shell subject to a concentrated load and to perform linear buckling analysis~\cite{Paulose2012ThermalShells}, which will be the topics for the following discussions. 
\subsection{Re-Deriving Some Established Results Using the Rescaling Transformation} 
We first demonstrate the convenience of the rescaling transformation by deriving the buckling load of orthotropic cylinders and plates from the corresponding isotropic expressions. 
Our results are consistent with the established expressions in literature. 
\subsubsection{Long Cylindrical Shells} \label{subsubsec: cylinders, buckling load} 
\paragraph{Edge Load.} \label{par: cylinders, buckling load, edge} 
By ``edge load'' we mean the load applied at the ends of an open cylindrical shell; it has units of pressure. 
Paschero and Hyer have observed the curious fact that the critical edge load of an orthotropic cylinder, when the real in-plane shear modulus is large enough (so that shear deformations are negligible), is exactly the classical buckling load of an isotropic cylinder with elastic constants $E_\eff$ and $\upsilon_\eff$~\cite{Paschero2009Ortho_Cylinders}. 
The rescaling transformation provides an explanation for this fact. 
The isotropic critical axial stress is in this case~\cite{Orange_Bible} 
\begin{equation} 
\sigma_\text{c, iso}^{11} = \frac{E}{ \sqrt{ 3(1 - \upsilon^2) } }\frac{t}{R}. 
\label{eqn: critical axial stress of isotropic cylinder} 
\end{equation} 
Since an orthotropic cylinder can be treated effectively as isotropic with a modified radius, we can use the same formula to write the \emph{rescaled} orthotropic critical stress: 
\begin{equation} 
\sigma_\text{c, ortho}^{1'1'} = \frac{E_\eff}{ \sqrt{ 3(1 - \upsilon_\eff^2) } }\frac{t}{R'}. 
\label{eqn: rescaled critical axial stress of orthotropic cylinder} 
\end{equation} 
Now recall that $\sigma^{1'1'} = \frac{1}{ \sqrt[4]{\lambda} }\sigma^{11}$, and $R' = \sqrt[4]{\lambda}R$. 
Substituting these into the above expression will yield the desired result 
\begin{equation} 
\sigma_\text{c, ortho}^{11} = \frac{E_\eff}{ \sqrt{ 3(1 - \upsilon_\eff^2) } }\frac{t}{R}. 
\label{eqn: critical axial stress of orthotropic cylinder} 
\end{equation}

\paragraph{Surface Load.} \label{par: cylinders, buckling load, surface} 
In this case, a uniform pressure is applied at the outer surface of an open cylindrical shell. 
The isotropic critical circumferential stress is known to be~\cite{Orange_Bible} 
\begin{equation} 
\sigma_\text{c, iso}^{22}t =      \frac{D}{R^2}\left(n^2 - 1         \right) 
                           \equiv \frac{D}{R^2}\left(n^2 - n_{\min}^2\right), 
\label{eqn: critical circumferential stress of isotropic cylinder} 
\end{equation} 
where $n$ is the number of half-waves in the circumferential direction. 
To obtain the orthotropic critical stress, we again substitute into the above expression the effective elastic constants and the rescaled quantities: 
\begin{align} 
\begin{split} 
    \sigma_\text{c, ortho}^{2'2'}t 
& = \frac{D'}{ {R'}^2 }\left( 
    { n'        }^2 
  - { n'_{\min} }^2 
    \right) 
    \\[0.25 em] 
    \sqrt[4]{\lambda}\sigma_\text{c, ortho}^{22}t 
& = \frac{\sqrt{\lambda}D_\theta}{\sqrt{\lambda}R^2}\left( 
    \sqrt[4]{\lambda}n^2 
  - \sqrt[4]{\lambda} 
    \right); 
\end{split} 
\label{eqn: rescaled critical circumferential stress of orthotropic cylinder} 
\end{align} 
$n' = \frac{R'}{R}\frac{y}{y'}n = \sqrt[8]{\lambda}n$ is the rescaled half-wave number (see \Eqref{eqn: Fourier coefficients for w, rescaled}). 
That it is not integral and related to the anisotropy parameter $\lambda$ arises from the following fact. 
Although distances and radii of curvature have the same dimension, the former are related to the square root of the metric, while the latter get rescaled in the same way as the metric since both the extrinsic curvature tensor and the metric tensor are rank-two. 
Cancelling all factors involving $\lambda$, we get 
\begin{equation} 
\sigma_\text{c, ortho}^{22}t = \frac{D_\theta}{R^2}\left(n^2 - 1\right), 
\label{eqn: critical circumferential stress of orthotropic cylinder} 
\end{equation} 
which is consistent with the result by Wang et al.~\cite{Wang2006Ortho_Cylinders}. 
\subsubsection{Plates} \label{subsubsec: plates, buckling load} 
We consider a rectangular orthotropic plate which is subject to in-plane compressive forces.
Its edges are aligned with the material orthotropic directions; the dimensions along the $x^1$ and $x^2$ directions are $a$ and $b$ respectively. 
The edges of the plate are simply supported; in other words, bending moments shall vanish along the edges which are held fixed but allowed to rotate during a deformation event (see \Eqsref{eqn: transforming the BCs}).
The plate is subjected to a uniform compression along the $x^1$ direction via edge loads of size $\tau$ per unit length acting upon the two edges perpendicular to the $x^1$ axis.
Force balance at equilibrium dictates a resulting compressive prestress $\sigma^{11} = \tau/t$.
We assume that shear deformations are negligible. 
In this case, the rescaling transformation maps the orthotropic plate with parameters 
$\left\{E_1, \upsilon_{21}, \lambda; a, b; \sigma^{11}\right\}$ to an isotropic plate with parameters
$\left\{E_\eff, \upsilon_\eff;
a' = \frac{a}{ \sqrt[8]{\lambda} }, 
b' = \sqrt[8]{\lambda}b; 
\sigma^{1'1'} 
\right\}$.
It should be pointed out that the orthotropic boundary conditions also become effectively isotropic, i.e., 
\begin{widetext} 
\begin{equation} 
\begin{cases} 
              \left. 
              u_3 
              \right|_{ \substack{x = 0, a \\ y = 0, b} } 
            = 0, 
              \\[1 em] 
\displaystyle \left. 
              \left(\pdfrac[2]{u_3}{x} + \upsilon_{12}\pdfrac[2]{u_3}{y}\right) 
              \right|_{x = 0, a} 
            = 0, 
              \\[1 em] 
\displaystyle \left. 
              \left(\pdfrac[2]{u_3}{y} + \upsilon_{21}\pdfrac[2]{u_3}{x}\right) 
              \right|_{y = 0, b} 
            = 0, 
\end{cases} 
\quad \mapsto \quad 
\begin{cases} 
              \left. 
              u_3 
              \right|_{ \substack{x' = 0, a' \\ y' = 0, b'} } 
            = 0, 
              \\[1 em] 
\displaystyle \left. 
              \left(\pdfrac[2]{u_3}{x'} + \upsilon_\eff\pdfrac[2]{u_3}{y'}\right) 
              \right|_{x' = 0, a'} 
            = 0, 
              \\[1 em] 
\displaystyle \left. 
              \left(\pdfrac[2]{u_3}{y'} + \upsilon_\eff\pdfrac[2]{u_3}{x'}\right) 
              \right|_{y' = 0, b'} 
            = 0. 
\end{cases} 
\label{eqn: transforming the BCs} 
\end{equation} 
\end{widetext} 
Therefore, the transformation only affects the way how quantities get ``measured" but does not change the system physically.

The resulting deformations manifest themselves as elastic waves. 
These waves are subject to the boundary conditions, \Eqsref{eqn: transforming the BCs}, and hence take the form 
$A_{mn}\sin\left(\frac{m\pi x}{a}\right)\sin\left(\frac{n\pi y}{b}\right)$, where $A_{mn}$ is the wave amplitude, and $m$ ($n$) denotes the number of half-waves propagating along the horizontal (vertical) direction. 
Because $x$ and $a$ ($y$ and $b$) rescale in the same way, $m' = m$ ($n' = n$), i.e., the half-wave numbers are invariant in this case (cf.\ \Eqsref{eqn: rescaled critical circumferential stress of orthotropic cylinder}). 
 
For an isotropic plate, the intensity of the load that gives rise to waves of a particular $(m, n)$ is given in~\Refref{Orange_Bible}: 
\begin{equation} 
\sigma_\text{iso}^{11}t = \frac{\pi^2D}{b^2}\left(\frac{m b}{a} + n^2\frac{a}{m b}\right)^2. 
\label{eqn: stresses of isotropic plates} 
\end{equation} 
The corresponding orthotropic stresses are hence 
\refstepcounter{equation} \label{eqn: stresses of orthotropic plates} 
\begin{align*} 
\sigma_\text{ortho}^{1'1'}t & = \frac{\pi^2D'}{b'^2}\left(\frac{m b'}{a'} + n^2\frac{a'}{m b'}\right)^2 
                                \tag{\theequation, a} 
                                \label{eqn: rescaled stresses of orthotropic plates} 
                                \\[0.25 em] 
\sigma_\text{ortho}^{1 1 }t & = \frac{\pi^2}  {b ^2}\left[ 
                                 D_1   \left(\frac{m b}{a}\right)^2 
                              + 2D' n^2 
                              +  D_2n^4\left(\frac{a}{m b}\right)^2 
                                \right], 
                                \tag{\theequation, b} 
                                \label{eqn: real stresses of orthotropic plates} 
\end{align*} 
which agrees with the known expression in the literature~\cite{Orange_Bible}. 
Its global minimum, with respect to the half-wave numbers, is the critical stress at which the plate buckles out of the plane.\footnote{
If the real shear modulus $G_{12}$ deviates much from the Huber form, to obtain the orthotropic stress, we can simply replace $D'$ in \Eqref{eqn: real stresses of orthotropic plates} with $H = \frac{G_{12}t^3}{6} + D_1\upsilon_{12}$, the real bending stiffness that penalizes twisting deformations (which reduces to $D'$ when assuming the Huber form). 
} 
\par 
In contrast to plates and singly-curved cylindrical shells with orthotropy, few exact results exist for the mechanical response of doubly-curved orthotropic shells. As a concrete application of our mapping, we next show that patches of orthotropic spheroidal shells transform locally to isotropic spheroidal shells with a different geometry, and use this mapping to derive new results for the indentation stiffness and buckling load of general orthotropic spheroidal shells. 
\subsection{Indentation Stiffness of Orthotropic Spheroidal Shells} \label{subsec: the local indentation stiffness}

The indentation assay---measuring the response of a structure to a point force---is commonly used to gauge the material properties of biological~\cite{Arnoldi2000,Smith2000,de_Pablo2003,Ivanovska2004,Deng2011,Schaap2006,Zhang2021}, as well as synthetic~\cite{Gordon2004,Zoldesi2008} shell structures. It also serves as a quantifier of shell stiffness and its relationships with geometry, pressure, and material properties which reveals the fundamental mechanisms underlying geometric rigidity~\cite{Vaziri2008,VellaSph,Vella2012Spheroid,Lazarus2012a,Wenqian2021}. Our mapping enables us to calculate the linear indentation response of orthotropic spheroidal and cylindrical shells under pressure, by making use of known results for isotropic shells~\cite{Wenqian2021}.

We consider an orthotropic spheroid, as depicted in \Figref{fig: orthotropic structures}, with a concentrated load exerted at the point $O$ on its equator. 
\subsubsection{The Zero-Pressure Case} \label{subsubsec: the zero-pressure stiffness} 
\paragraph{General Doubly-Curved Shells.} \label{par: general shells, stiffness, pressureless} 
For this simple case, $\sigma_0^{\alpha'\beta'} = 0$, and an extra term, $-F\delta(x^1)\delta(x^2)$, needs to be included on the right-hand side of \Eqref{eqn: the linearized EOE} to model the concentrated load at the origin, where $F$ denotes the load strength, and $\delta(x)$ is the Dirac delta function. 
Note that because of the scaling property of the delta function, $\delta(ax) = \frac{1}{ \abs{a} }\delta(x)$, the load strength does not need rescaling, i.e., 
$
  F \delta     (x^1         )\delta     (x^2         ) 
= F \delta\left(x^{1'}\right)\delta\left(x^{2'}\right) 
\equiv 
  F'\delta\left(x^{1'}\right)\delta\left(x^{2'}\right) 
$. 
\par 
The indentation stiffness is defined as 
\begin{equation} 
k \coloneqq -\frac{F }{ u_3   (0, 0) } 
          = -\frac{F'}{ u_{3'}(0, 0) }, 
\label{eqn: the def of indentation stiffness} 
\end{equation}
where $u_{3}(0,0)$ is the transverse displacement of the shell at the origin in response to the indentation load.
As shown in \Refref{Wenqian2021}, the inverse of the indentation stiffness \emph{at zero pressure} (denoted by $k^0$) is given by the following definite integral: 
\begin{equation} 
  \frac{1}{k^0} 
= \frac{1}{4\pi^2}\int_{\mathbb{R}^2} 
  \frac{ Q\, \df{Q}\, \df{\varphi} } 
  {\displaystyle 
  D' Q^4 
+ Y'\left( 
  \frac{1}{ R_{2'} }\cos^2{\varphi} 
+ \frac{1}{ R_{1'} }\sin^2{\varphi} 
  \right)^2 
  }, 
\label{eqn: zero-pressure stiffness integral} 
\end{equation} 
where the integration variables $Q$ and $\varphi$ are related to wavevectors, $\mathbf{q} = \VEC{ q^{1'}, q^{2'} }$, in the following way: 
$q^{1'} = \frac{1}{ \sqrt[4]{\lambda} }Q\cos{\varphi}$ and 
$q^{2'} =           \sqrt[4]{\lambda}  Q\sin{\varphi}$. 
The fact that $Q^2 = \sqrt{\lambda}\left(q^{1'}\right)^2 + \frac{1}{ \sqrt{\lambda} }\left(q^{2'}\right)^2$ implicitly reflects that the metric of the rescaled Fourier space is non-Euclidean, resulting from the original material orthotropy. 
Evaluating the integral in \Eqref{eqn: zero-pressure stiffness integral} gives 
\begin{align} 
    k^0 = 8\sqrt{D' Y'}\sqrt{K'} 
& = \frac{4 E_\eff      t^2}{ \sqrt{ 3\left(1 - \upsilon_\eff^2           \right) } }\frac{1}{ \sqrt{R_1R_2} } 
    \nonumber \\[0.25 em] 
& \coloneqq 
    \frac{4\sqrt{E_1E_2}t^2}{ \sqrt{ 3\left(1 - \upsilon_{12}\upsilon_{21}\right) } }\frac{ \sqrt{1 - \beta_0} }{R_2}, 
\label{eqn: zero-pressure stiffness} 
\end{align} 
where in the second line, we used the definition of the effective elastic constants, $E_\eff$ and $\upsilon_\eff$. 
The parameter $\beta_0 \coloneqq 1 - \frac{R_2}{R_1}$ characterizes the asphericity of a given spheroid; for example, $\beta_0 = 0$ ($R_1 = R_2$) corresponds to a sphere, while $\beta_0 = 1$ ($R_1 \to +\infty$) a cylinder. 
\Eqnref{eqn: zero-pressure stiffness} clearly shows the separate contributions of geometry and material anisotropy to the indentation stiffness. 
As in the isotropic case, the Gaussian curvature, $K = K' = 1/(R_1 R_2)$, is still the dominant geometrical quantity that governs shell stiffness at zero pressure~\cite{Vella2012Spheroid, Wenqian2021}. 
Heuristically, we could have anticipated this $K$ dependence based on the fact that $K$ is invariant under our rescaling transformation (\Eqref{eqn: invariance of Gaussian curvature}), and therefore captures the geometric rigidity independently of how the rescaling is performed. 
\par 
Just as the geometric contribution is captured by the geometric mean of the two curvatures, effects of material anisotropy also come in the form of geometric means, $E_\eff = \sqrt{E_1E_2}$ and 
$\upsilon_\eff = \sqrt{ \upsilon_{12}\upsilon_{21} }$. 
These geometric-mean dependencies are consistent with the requirement of invariance of the indentation stiffness under coordinate transformations. 
Consider the equator of an orthotropic sphere. 
We call the local polar (meridional) and azimuthal (zonal) direction the first and the second direction, respectively, i.e., $\theta \equiv x^1$ and $\phi \equiv x^2$. 
Assume that the shell is strengthened along the first direction, i.e., $E_1 > E_2$. 
We now rotate our local coordinate system clockwise by ninety degrees, so that $x^1 \mapsto -x^2$, and $x^2 \mapsto x^1$. 
The rotation leaves us with the same spherical shell locally but with the second direction strengthened. 
We can infer two conclusions from this simple argument. 
First, any \emph{local} elastic property around the equator of an orthotropic sphere should exhibit an exchange symmetry: Interchanging directions $1$ and $2$ does not make a difference. 
Second, if material anisotropy and geometry affect shell elasticity locally separately from one another, then for shells of \emph{any} type, their local elastic properties should depend on combinations of elastic constants which are invariant under the interchange $1 \leftrightarrow 2$. 
\par 
More generally, local elastic properties should be functions of invariant quantities constructed from the corresponding tensors. 
In our system, such examples are furnished by the Gaussian curvature (the square root of the extrinsic curvature tensor's \emph{determinant}) as well as the combination $\frac{ 1 - \upsilon_{12}\upsilon_{21} }{E_1E_2}$, which happens to be the determinant of the stiffness matrix $C_{ [\alpha\beta][\gamma\delta] }$ (\Eqref{eqn: the generalized Hooke's law}), if we assume that the deformation is axisymmetrical, i.e., ignoring $G_{12}$. We can to some extent rule out the \emph{trace} of these tensors as the invariant setting the stiffness, based on the fact that these traces are not invariant under the rescaling transformation (cf.\ the local mean curvature, \Eqref{eqn: mean curvature under rescaling}). 
\par 
For general ellipsoidal shells of revolution, material properties are usually different along the polar and azimuthal direction, which are also the principal directions of such shell surfaces~\cite{do_Carmo_DiffGeo}. 
Therefore, according to the shallow-shell theory, for these shells, \Eqref{eqn: zero-pressure stiffness} can be applied almost globally, except at the two poles, where the material-orthotropy pattern becomes curvilinear. 
Nevertheless, if we think of the local indentation stiffness as a function of positions on the shell surface and consider only small deformations, we expect that taking the analytical continuation of the function to the poles will imply that \Eqref{eqn: zero-pressure stiffness} can be still valid there. 
An explicit calculation of the zero-pressure indentation stiffness at the poles confirms this expectation for spherical shells (see  \Appref{appendix: pole calculations} for calculation details and comparison to simulations). 
\paragraph{Long Cylindrical Shells.} \label{par: cylinders, stiffness, pressureless} 
As for their isotropic counterparts~\cite{Wenqian2021}, the case of long orthotropic cylinders also requires special attention.
In contrast to doubly-curved shells, cylinders admit nearly-isometric deformations which are not accurately captured by the two-dimensional Fourier transform applied to a shallow shell section as used in \Eqref{eqn: zero-pressure stiffness integral}.
Instead, the transverse deformation field along the entire circumferential direction must be described using a Fourier series; this approach was used by Yuan to describe the indentation of  isotropic cylinders~\cite{Yuan1946_Cylinders}.
We apply the rescaling transformation to Yuan's analysis and accordingly obtain the following expression for the zero-pressure stiffness of orthotropic cylinders (see \Appref{appendix: Yuan's analysis} for details): 
\def\cyl { \mathrm{cyl } } 
\def\even{ \mathrm{even} } 
\begin{widetext} 
\begin{equation} 
  k_\cyl^0(\lambda) 
\approx 
  \frac{1}{ \sqrt[4]{\lambda} } 
  \frac{2\pi}{ 3\sqrt{2}\left(1 - \upsilon_\eff^2\right) } 
  \frac{E_\eff t^3}{R^2} 
  \left( 
  \sum_{n = 1}^\infty \frac{1}{n^3} 
  \frac{ \sqrt{1 + \Xi_n} }{\Xi_n} 
  \right)^{-1}, 
\label{eqn: zero-pressure k, long cylinders} 
\end{equation} 
where 
$
\Xi_n^2 \coloneqq 1 
                + \frac{ 3\left(1 - \upsilon_\eff^2\right) }{4n^4} 
                  \left(\frac{R}{t}\right)^2 
$. 
The dependence of the stiffness expression on the cylinder's thickness and radius in the thin-shell limit ($R/t \gg 1$) is obtained by keeping the leading term of the series in \Eqref{eqn: zero-pressure k, long cylinders}, which dominates when $\frac{R}{t}$ is sufficiently large: 
\begin{equation} 
        k_\cyl^0(\lambda) 
\approx \frac{1}{ \sqrt[4]{\lambda} } 
        \frac{2\pi}{ 3\sqrt{2}\left(1 - \upsilon_\eff^2\right) } 
        \frac{E_\eff t^3}{R^2} 
        \sqrt{\Xi_1} 
\approx \frac{1}{ \sqrt[4]{\lambda} } 
        \frac{\pi}{ \left[3\left(1 - \upsilon_\eff^2\right)\right]^\frac{3}{4} } 
        \frac{ E_\eff t^\frac{5}{2} }{ R^\frac{3}{2} }. 
\label{eqn: zero-pressure k, long cylinders, scaling form} 
\end{equation} 
\end{widetext} 
For isotropic ($\lambda = 1$) cylinders with a negligible Poisson's ratio ($\upsilon \approx 0$), \Eqref{eqn: zero-pressure k, long cylinders, scaling form} becomes 
\begin{equation} 
        k_\cyl^0(\lambda = 1) 
\approx 1.38\frac{ Et^\frac{5}{2} }{ R^\frac{3}{2} }. 
\label{eqn: zero-pressure k, long isotropic cylinders, scaling form} 
\end{equation} 
\Eqnref{eqn: zero-pressure k, long isotropic cylinders, scaling form} matches exactly, including the order-one prefactor, with the expression obtained by de Pablo et al.~\cite{de_Pablo2003}. 
\par 
From \Eqref{eqn: zero-pressure k, long cylinders}, we observe that the zero-pressure indentation stiffness for long cylinders depends on the anisotropy parameter $\lambda$ both implicitly (through the anisotropic elastic constants absorbed into $E_\eff$ and $\upsilon_\eff$) and explicitly (in the $1/\sqrt[4]\lambda$ factor), unlike the stiffness of orthotropic doubly-curved shells whose $\lambda$-dependence is purely implicit (see \Eqref{eqn: zero-pressure stiffness}). 
The explicit $\lambda$-dependence breaks the aforementioned local exchange symmetry and is a consequence of the fact that open cylinders can deform isometrically (see \textbf{\ref{par: cylinders, stiffness, pressurized}}). 
\par 
\Figureref{fig: zero-pressure indentation stiffness, spheroids and cylinders}  compares the theoretical predictions for zero-pressure indentation stiffness of orthotropic spheroids (curves) to the output of finite-element simulations (symbols).
For each geometry, we report a non-dimensionalized stiffness $\tilde{k}$ obtained by dividing $k^0$ by the zero-pressure stiffness of a spherical (for $\beta_0 < 1$) or cylindrical (for $\beta_0 = 1$) shell with the same equatorial radius $R = R_2$ and  \emph{isotropic} elasticity governed by  $\{E = E_1, \upsilon = \upsilon_\eff\}$. 
This stiffness scale was chosen to show both the implicit and explicit $\lambda$-dependencies as well as the dependence on Gaussian curvature for doubly-curved shells (\Eqref{eqn: zero-pressure stiffness}): The Gaussian curvature can be written as $K = \frac{1 - \beta_0}{R_2^2}$, so the zero-pressure stiffness of a doubly-curved shell with a lower $\beta_0$ is larger ($R_2$ is fixed).

Theoretical predictions lie within a few percent of finite-element measurements for all parameter values in \Figref{fig: zero-pressure indentation stiffness, spheroids and cylinders}, verifying that our rescaling transformation provides accurate results for indentation calculations.
The largest discrepancy between theory and simulation is around $8\%$ for cylinders with $\lambda < 1$; this mismatch likely stems from the fact that \Eqref{eqn: zero-pressure k, long cylinders, scaling form} omits higher-order terms.
Despite the simplification, the expression accurately captures the explicit $\lambda$-dependence of the zero-pressure cylinder stiffness, which follows a different power-law relationship compared to doubly-curved shells as seen in the insets to \Figref{fig: zero-pressure indentation stiffness, spheroids and cylinders}.
From the top inset, we can see that the indentation stiffness of doubly-curved shells all scales as $\frac{1}{ \sqrt{\lambda} }$, which shows the dependence on $E_\eff$ (since $E_\eff = E_1 / \sqrt{\lambda}$). 
In contrast, the stiffness of long cylinders has a $\lambda$-dependence given by $\frac{1}{ \sqrt[4]{\lambda^3} }$; this is a combination of the same $E_\eff$ dependence and the explicit $\frac{1}{ \sqrt[4]{\lambda} }$ factor in \Eqref{eqn: zero-pressure k, long cylinders}. 

\begin{figure}[t] 
\centering 
\includegraphics[width = 0.48 \textwidth]{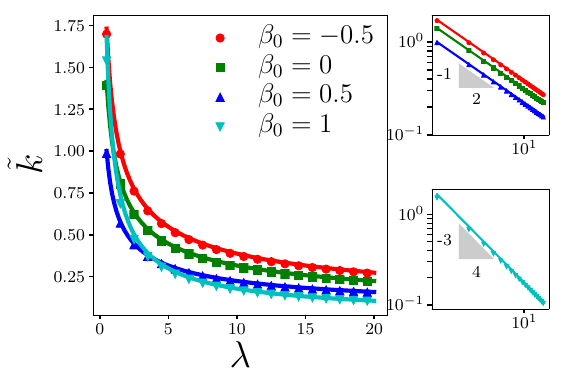} 
\caption{
Zero-pressure indentation stiffness of four different types of orthotropic shells with varying values of the anisotropy parameter $\lambda$. 
Symbols denote data obtained from finite-element simulations as detailed in \Appref{app: comsol}. 
Solid curves correspond to the analytical expressions \Eqsref{eqn: zero-pressure stiffness} and \eqref{eqn: zero-pressure k, long cylinders, scaling form}. 
The vertical axis reports the stiffness scaled by $
\frac{4E_1 t^2}{ \sqrt{ 3\left(1 - \upsilon_\eff^2\right) } }\frac{1}{R_2} 
$ for doubly-curved shells ($\beta_0 = 0, \pm0.5$), and by  
$
\frac{\pi}{ \left[3\left(1 - \upsilon_\eff^2\right)\right]^\frac{3}{4} } 
\frac{ E_1 t^\frac{5}{2} }{ R^\frac{3}{2} } 
$ for cylinders ($\beta_0 = 1$). 
The insets show the same data on logarithmic axes. 
} 
\label{fig: zero-pressure indentation stiffness, spheroids and cylinders} 
\end{figure} 
\subsubsection{The Pressurized Case} \label{subsubsec: stiffness of pressurized shells} 

We now consider the indentation stiffness of closed orthotropic shells subjected to a uniform pressure. This situation is relevant to biological shell-like structures, which often experience high turgor pressures; varying the pressure also provides a route to modifying the shape and stiffness of artificial shells~\cite{Datta2012,Herrmann2019}. 

In the absence of indenting forces, the pressurized shell deforms from its original shape to attain a new equilibrium in which in-plane stresses balance the transverse loads due to the pressure.
The indentation forces and deflections are then calculated with reference to this prestressed state.
For thin shells, the indentation response is still a local property of the geometry, elasticity, and prestresses in the vicinity of the indentation point, and a shallow-shell description of the local response will suffice to calculate the indentation stiffness.
However, the prestressed state itself depends on the \emph{global} shell shape---it is not determined solely by local properties~\cite{Timoshenko_Bible}. 
For thin spheroidal and cylindrical orthotropic shells, these prestress configurations in response to a uniform pressure are known as a function of pressure and global geometry~\cite{Orange_Bible}, and are independent of the elastic properties of the shell as long as the deformations in response to the pressure are small. We will use these prior results as inputs to our rescaled theory, which we then use to calculate the indentation stiffness as a function of geometry and pressure.

\paragraph{General Spheroids.} \label{par: spheroids, stiffness, pressurized} 
\def\sph{ \mathrm{sph} } 
Spheroids are ellipsoids of revolution. 
We are interested in the local indentation stiffness around a spheroid's equator. 
Following~\Refref{Wenqian2021}, we use $\beta_0 \coloneqq 1 - \frac{R_2}{R_1}$ to characterize the asphericity of a spheroid, where $R_2$ is the radius of its equator, and $\frac{1}{R_1}$ is the local principal curvature along the meridional direction for points on the equator. 
In the vicinity of the equator, the prestress components are given by 
$\sigma_0^{11}t = \frac{1}{2}pR_2             $, 
$\sigma_0^{22}t = \frac{1}{2}pR_2(1 + \beta_0)$ and 
$\sigma_0^{12}t = 0                           $~\cite{Orange_Bible}, where $p$ denotes the uniform pressure to which the spheroid is subject.\footnote{
We would like to mention that an orthotropic spheroid shares the same prestress as the corresponding isotropic one with the same geometry \emph{only} on regions that are far away from the two poles~\cite{Steele1965, Reissner1958}. 
} 
The sign convention for the pressure is that a positive (negative) $p$ means an internal (external) pressure. 
\par 
Following the same procedure as the zero-pressure case, we obtain the inverse of the indentation stiffness which is now a function of three parameters, namely, the scaled pressure $\p(\lambda) = \frac{pR_2^2}{ 4\sqrt{ D'(\lambda)Y'(\lambda) } }$, the asphericity $\beta_0$ as well as the anisotropy parameter $\lambda$ \emph{explicitly}:
\begin{widetext} 
\begin{equation} 
  \frac{1}{ k(\p(\lambda), \beta_0, \lambda) } 
= \frac{1}{8\pi^2}\sqrt{ \frac{R_{2'}^2}{D' Y'} }\int_0^{2\pi} \df{\varphi}\, 
  \int_0^{+\infty} 
  \frac{ \df{u} } 
  { 
  u^2 
+ 2\p\left(1 + \beta'_\lambda\sin^2{\varphi}\right)  u 
+    \left(1 - \beta'        \sin^2{\varphi}\right)^2 
  }, 
\label{eqn: pressurized k, spheroids} 
\end{equation} 
where 
$\beta'         \coloneqq 1 - \sqrt{\lambda}(1 - \beta_0)$ and 
$\beta'_\lambda \coloneqq 2\sqrt{\lambda} - 2 + \beta'   $ appear to couple the geometry and the material anisotropies. 
Nonetheless, it turns out that these explicit $\lambda$-dependences are spurious, as we will now demonstrate. 
The double integral in \Eqref{eqn: pressurized k, spheroids} can be evaluated in the following closed form (see \Appref{appendix: evaluating integrals} for details): 
\begin{equation} 
  \frac{1}{ k(\p(\lambda), \beta_0) } 
= \frac{1}{2\pi}\sqrt{ \frac{R_2^2}{D' Y'} } 
  \frac{1}{ \sqrt{1 - \beta_0} }\frac{1}{ \sqrt{ (1 - \p)(1 + \alpha\p) } } 
  \EllipticF 
  { \frac{1}{2}\arccos{\p} } 
  { 
- \frac{2(1 - \alpha)\p}{ (1 - \p)(1 + \alpha\p) } 
  }, 
\label{eqn: pressurized k, spheroids, the closed form} 
\end{equation} 
where $\alpha \coloneqq \frac{1 + \beta'_\lambda}{1 - \beta'} = \frac{1 + \beta_0}{1 - \beta_0}$ is independent of $\lambda$, and $\EllipticF{\vartheta}{C^2}$ denotes the incomplete elliptic integral of the first kind: 
\begin{equation} 
\EllipticF{\vartheta}{C^2} \coloneqq \int_0^\vartheta \frac{ \df{\varphi} }{ \sqrt{ 1 - C^2\sin^2{\varphi} } }. 
\label{eqn: EllipticF} 
\end{equation} 
\par 
\Eqnref{eqn: pressurized k, spheroids, the closed form} is the primary result of this work: by applying the rescaling transformation, we have obtained a closed-form expression for the equatorial indentation stiffness of pressurized orthotropic spheroids, provided that the material anisotropy directions align with the latitudinal and longitudinal directions as shown in \Figref{fig: orthotropic structures}. As a consistency check, setting $\beta_0 = 0$ ($R_1 = R_2 = R$) and $\lambda = 1$ ($E_\eff = E$ and $\upsilon_\eff = \upsilon$) reduces \Eqref{eqn: pressurized k, spheroids, the closed form} to, after taking the inverse, 
\begin{equation} 
  k(\p(1), 0) 
= \frac{ 2\pi\sqrt{D Y} }{R} 
  \frac{ \sqrt{1 - \eta^2} }{ \displaystyle \EllipticF{ \frac{1}{2}\arccos{\eta} }{0} } 
= \frac{ 8\sqrt{D Y} }{R} 
  \frac{ \sqrt{1 - \eta^2} }{ \displaystyle 1 - \frac{2}{\pi}\arcsin{\eta} }, 
\label{eqn: pressurized k, isotropic spheres} 
\end{equation} 
which recovers the established result of the indentation stiffness of pressurized isotropic spherical shells~\cite{VellaSph, Paulose2012ThermalShells}. 
\end{widetext} 

\Eqnref{eqn: pressurized k, spheroids, the closed form} demonstrates that the indentation stiffness depends on the anisotropy parameter $\lambda$ only \emph{implicitly} via the coupling constants $D' = \sqrt{D_1D_2}$ and $Y' = E_\eff t \equiv \sqrt{E_1E_2}t$. 
In other words, the \emph{only} effect of material anisotropy is modifying the elastic constants. 
As a consequence, our previous analysis on the behavior of the stiffness integral in different parameter regimes~\cite{Wenqian2021} should carry over to the orthotropic case.
In particular, it was established in \Refref{Wenqian2021} that at high pressures, the geometry of the spheroid becomes less relevant and instead the indentation response is dominated by a new length scale---the \emph{radius of distensile curvature}, defined as $$\mathcal R \equiv \frac{1}{p}\sqrt{ \det\left(\sigma_0^{\alpha\beta}t\right) } = R_2 \sqrt{1+\beta_0};$$
the indentation stiffness for arbitrary isotropic ellipsoids at high pressure approaches that of a sphere of radius $\mathcal{R}$ and experiencing the same pressure.
For anisotropic spheroids, we expect the same behavior, provided the geometric-mean coupling constants $D'$ and $Y'$ are used to define the relevant pressure scale: upon defining a non-dimensionalized pressure $\eta_\mathcal{R} \equiv p\mathcal{R}^2/(4\sqrt{D'Y'})$, we expect the rescaled indentation stiffness $\tilde{k} \equiv k/\sqrt{ \frac{D' Y'}{ \mathcal{R}^2 } }$ for different shell geometries to approach a single curve when $\eta_\mathcal{R} \gg 1$.

This expectation is confirmed in \Figref{fig: indentation stiffness, pressurized spheres and cylinders}, which reports simulation data (symbols) and theoretical predictions (solid curves) for the indentation stiffness of pressurized shells as a function of pressure over a range of geometry and anisotropy values.
The data have been nondimensionalized  using scales related to the radius of distensile curvature $\mathcal{R}$ and the geometric-mean elastic constants $D'$ and $Y'$.
Using this rescaling, indentation stiffnesses measured from simulations with different material anisotropies collapse onto curves that depend only on the geometry parameter $\beta_0$.
The data collapse and agreement with the solid curves for $\beta_0 \in \{-0.778,0,0.75, 0.96\}$ validate our prediction for the indentation stiffness of pressurized orthotropic doubly-curved shells, \Eqref{eqn: pressurized k, spheroids, the closed form}. 
Theoretical curves and simulation data for different shell geometries  converge in the limit $\eta_{\mathcal R} \gg 1$, indicating that the distensile curvature and the pressure-induced prestresses fully determine the indentation response for shells with large internal pressures as anticipated by the behavior of pressurized isotropic shells (\Refref{Wenqian2021}). 
\paragraph{Long Cylindrical Shells.} \label{par: cylinders, stiffness, pressurized} 
We had previously mentioned that shallow-shell theory failed to capture the indentation stiffness of cylinders at zero pressure, which instead required a different analysis (Section \ref{par: cylinders, stiffness, pressureless} and \Appref{appendix: Yuan's analysis}).
However, we expect that the shallow-shell approach again becomes accurate for cylinders at large enough internal pressures~\cite{Wenqian2021}.
At finite internal pressure, the extent of the indentation deformation along the circumferential direction of the cylinders is restricted by a length scale of order $\sqrt{D_2/(p R_2)}$, which becomes much smaller than $R_2$ at high enough pressures.
In this regime, the shallow-shell analysis leading to \Eqref{eqn: pressurized k, spheroids, the closed form} for doubly-curved shells is also appropriate for cylinders.
It is possible to obtain the indentation stiffness of long pressurized cylinders directly from \Eqref{eqn: pressurized k, spheroids, the closed form} by taking the limit $\beta_0 \to 1^-$ and invoking \LH's rule repeatedly. 
As a more direct approach, we first impose that limit in \Eqref{eqn: pressurized k, spheroids} and then evaluate the resulting definite integral. 
By doing so, we get, after some calculations (see \Appref{appendix: evaluating integrals}), 
\begin{widetext} 
\begin{align} 
\begin{split} 
    \frac{1}{ k(\p(\lambda), 1) } 
  \equiv 
    \frac{1}{ k_\cyl( \p(\lambda) ) } 
& = \frac{1}{4\pi}\sqrt{ \frac{R^2}{D' Y'} } 
    \frac{1}{ \sqrt{\p} }\int_0^{+\infty} 
    \frac{ \df{x} }{ \sqrt{x^4 + 2\p x^2 + 1} } 
    \\[0.25 em] 
& = \frac{1}{4\pi}\sqrt{ \frac{R^2}{D' Y'} } 
    \frac{1}{ \sqrt{\p} }\EllipticK{ \frac{1}{2}(1 - \p) }, 
\end{split} 
\label{eqn: pressurized k, long cylinders} 
\end{align} 
where $R \equiv R_2$ is the radius of cylinders, and $\EllipticK{C^2}$ denotes the complete elliptic integral of the first kind: 
\begin{equation} 
\EllipticK{C^2} \coloneqq \EllipticF{ \frac{\pi}{2} }{C^2} 
                        = \int_0^\frac{\pi}{2} \frac{ \df{\varphi} }{ \sqrt{ 1 - C^2\sin^2{\varphi} } }. 
\label{eqn: EllipticK} 
\end{equation} 
\end{widetext} 
The fact that $\EllipticK{C^2}$ is analytic for $\abs{C} < 1$ indicates that when $0 < \p \ll 1$, $k_\cyl( \p(\lambda) ) \propto \sqrt{\p}$ for orthotropic cylinders, just like their isotropic counterparts~\cite{Wenqian2021}.

\begin{figure}[tb] 
\centering 
\includegraphics[width = 0.48\textwidth]{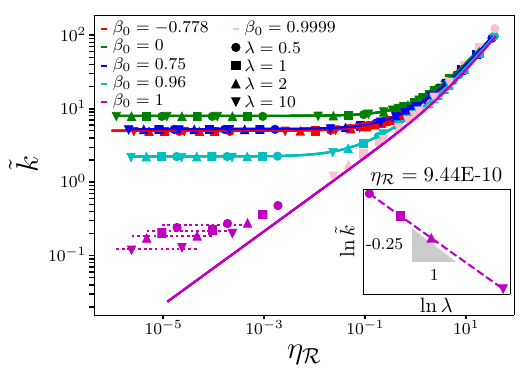} 
\caption{
Indentation stiffness of orthotropic shells of varying geometries ($\beta_0$ values) and degrees of anisotropy ($\lambda$ values), as a function of pressure. 
Symbols denote data obtained from finite-element indentation simulations as described in \Appref{app: comsol}. 
Solid curves correspond to the analytical expressions \Eqref{eqn: pressurized k, spheroids, the closed form} for $\beta_0 < 1$ and \Eqref{eqn: pressurized k, long cylinders} for $\beta_0 = 1$. 
Data are scaled using the stiffness scale $\sqrt{ \frac{D' Y'}{ \mathcal{R}^2} }$ and the pressure scale $\frac{ 4\sqrt{D' Y'} }{ \mathcal{R}^2}$ derived from the distensile curvature radius $\mathcal{R}$. 
Dotted horizontal lines indicate the scaled zero-pressure stiffness of orthotropic cylinders (\Eqref{eqn: zero-pressure k, long cylinders, scaling form}). 
Inset, indentation stiffness of cylinders simulated at a low pressure $\eta_{\mathcal R} = 9.44 \times 10^{-10}$. The dashed line indicates the power-law relationship $\tilde{k} \propto \frac{1}{\sqrt[4]{\lambda}}$. 
} 
\label{fig: indentation stiffness, pressurized spheres and cylinders} 
\end{figure}

To test the validity of our prediction for the pressurized cylinder stiffness, \Eqref{eqn: pressurized k, long cylinders}, we compared the expression to finite-element simulation measurements for nearly cylindrical spheroids with $\beta_0 = 0.9999$ and varying levels of material anisotropy in \Figref{fig: indentation stiffness, pressurized spheres and cylinders}. The highly elongated spheroidal geometry was used to avoid instability and convergence issues with simulating perfect cylinders at high pressures (see \Appref{subsubsec: spheroids in COMSOL} for details). We found that the nearly cylindrical shells closely follow \Eqref{eqn: pressurized k, long cylinders} (solid curve, $\beta_0 = 1$)  for high enough rescaled pressures ($\eta_\mathcal{R} \gtrsim 10^{-2}$).
We also simulated perfect cylinders ($\beta_0 = 1$, symbols) of finite length and rescaled pressure values below $10^{-2}$.
At these low pressures, the indentation stiffness deviates from the expression derived from shallow-shell theory, and instead approaches the zero-pressure expression evaluated using a Fourier series, \Eqref{eqn: zero-pressure k, long cylinders, scaling form} (dotted lines).   
As noted above, the zero-pressure stiffness of anisotropic cylinders retains an explicit $\lambda$-dependence beyond the implicit dependence through the combination of elastic moduli $D'Y'$; this dependence is made apparent by the fact that the cylinder data at low pressures and different anisotropy values no longer collapse onto each other in \Figref{fig: indentation stiffness, pressurized spheres and cylinders}.
The inset confirms that the residual $\lambda$-dependence of the rescaled indentation stiffness for cylinders at very low internal pressures follows the expectation $\tilde{k} \propto k^0_\text{cyl}/E_\text{eff} \propto \frac{1}{\sqrt[4]{\lambda}}$ from \Eqref{eqn: zero-pressure k, long cylinders}. 
The results for shells with $\beta_0 = 0.9999$ and $\beta_0 = 1$ show that the expressions \Eqref{eqn: zero-pressure k, long cylinders} and  \Eqref{eqn: pressurized k, long cylinders}, taken together, provide a nearly comprehensive analytical understanding of the indentation stiffness of pressurized orthotropic cylinders. 
 
\subsection{Buckling Load of Orthotropic Spheroids under Uniform Pressure} \label{subsec: the buckling load}

Shell buckling is a catastrophic failure mode of thin-walled structures, and its avoidance is of critical importance in engineering design~\cite{Timoshenko_Bible,Koiter_notes}.
Buckling is also one of the key mechanisms that give rise to a diversity of morphologies in nature, ranging from saddle-shaped leaves~\cite{Liang2019_Leaves} to the undulating shapes of animal tissues in diverse organs~\cite{Nelson2016}. 
In technology,  buckling can be exploited for actuation and shape control of soft capsules~\cite{Shim2012,Datta2012}, with a range of potential applications in, e.g., 4D printing and drug delivery~\cite{Lin2018_Buckling_Review}.
Although shell buckling is a nonlinear phenomenon, the buckling load of a shell can be predicted by linear stability analysis~\cite{Orange_Bible}. 
We now use our mapping to generate expressions for the critical buckling pressures of spheroids in parameter regimes for which buckling is driven by a linear instability in a region of local rectilinear orthotropy.

\subsubsection{General Spheroids} \label{subsubsec: spheroids, buckling load} 
\paragraph{External Buckling Pressure.} \label{par: spheroids, buckling load, external} 
When a curved shell buckles under a uniform pressure, it also becomes locally soft, i.e., its indentation stiffness vanishes, because of the emergence of an unstable mode, for which the integral in \Eqref{eqn: pressurized k, spheroids} diverges~\cite{Wenqian2021}. 
We can then obtain the \emph{local} buckling pressure around the equator of such shells by studying the zeros of $k(\p(\lambda), \beta_0)$ for a given $\beta_0$. 
We can read off the zeros directly from \Eqref{eqn: pressurized k, spheroids, the closed form} and hence acquire the nondimensionalized buckling pressure: 
\refstepcounter{equation} \label{eqn: the buckling pressure of orthotropic spheroids} 
\begin{equation} 
  \cp 
= \begin{cases} 
- 1, 
  & \text{for the oblate  } (\beta_0 \leq 0), 
  \\[0.25 em] 
  \displaystyle 
- \frac{1 - \beta_0}{1 + \beta_0}, 
  & \text{for the prolate } (\beta_0 > 0). 
\end{cases} 
\tag{\theequation, a} 
\label{eqn: the buckling pressure of orthotropic spheroids, scaled} 
\end{equation} 
Recall that the pressure scale used in \Eqref{eqn: pressurized k, spheroids, the closed form} is $p_\mathrm{sc} \coloneqq \frac{ 4\sqrt{D' Y'} }{R_2^2}$. 
The dimensionful buckling pressure is thus 
\begin{equation} 
  \CP 
\coloneqq 
  \cp p_\mathrm{sc} 
= \begin{cases} 
  \displaystyle 
- \frac{ 4\sqrt{D' Y'} }{R_2^2}, 
  & \text{for } \beta_0 \leq 0, 
  \\[1.25 em] 
  \displaystyle 
- \frac{ 4\sqrt{D' Y'} }{2R_1R_2 - R_2^2}, 
  & \text{for } \beta_0 > 0. 
\end{cases} 
\tag{\theequation, b} 
\label{eqn: the buckling pressure of orthotropic spheroids, dimensionful} 
\end{equation} 

As was the case with the indentation stiffness expression, the local buckling pressure of orthotropic spheroids (both prolate and oblate) is exactly that of the corresponding isotropic shells~\cite{Wenqian2021} with the same geometry and with geometric-mean elastic constants taking the place of the isotropic elasticity parameters. 
This fact again shows that the main effect of material anisotropy is to modify the elastic constants; the geometric contribution (radius dependence of the buckling pressure) is not affected. 
Our result is consistent with the established expression for the buckling pressure of spheroidal shells stiffened by reinforcements along the equatorial or longitudinal directions, which was also founded on the shallow-shell theory~\cite{Becker1968_Ship_Research}. 

As a special case, the local buckling pressure of an orthotropic sphere around its equator is given by setting $\beta_0 = 0$ above:  
\begin{equation} 
p_\text{c, sph} = -\frac{ 4\sqrt{D' Y'} }{R^2}, 
\label{eqn: the real buckling pressure of orthotropic spheres} 
\end{equation}
The buckling of orthotropic spheres was investigated computationally and experimentally in \Refref{Herrmann2019}. In that work, it was found that upon increasing the external pressure on an orthotropic sphere with material anisotropy aligned to the polar and azimuthal directions, buckling first occurred in the vicinity of the equator when  $\lambda \geq 1$ (i.e., when the stiffness $E_1$ along the polar direction is greater than the stiffness $E_2$ along the azimuthal direction). 
Consequently, our expression for the local buckling pressure at the equator provides a prediction for the \emph{global} buckling pressure when $\lambda > 1$.

\begin{figure}[tb] 
\centering 
\includegraphics[width = 0.48 \textwidth]{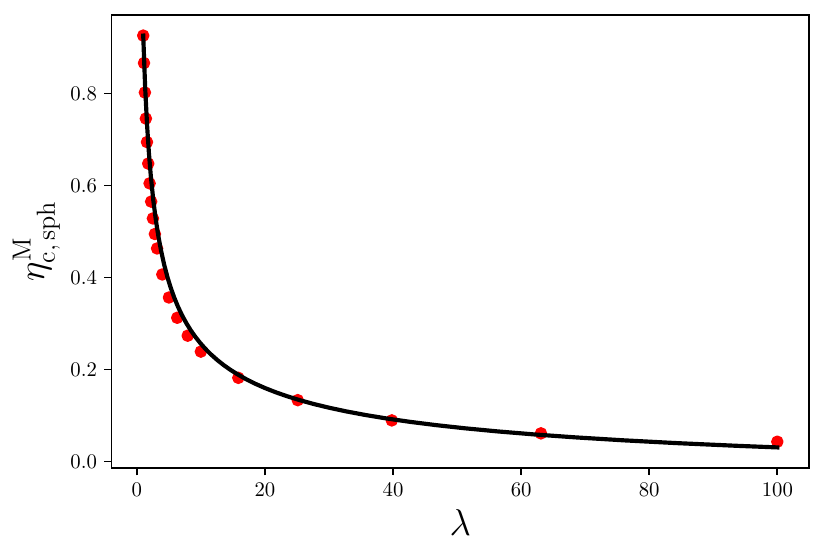} 
\caption{
The scaled global buckling pressure of orthotropic spheres as a function of the degree of material anisotropy $\lambda$. 
The spheres considered here have a larger Young's modulus along the polar direction, i.e., $\lambda \geq 1$ ($E_1  \geq  E_2$).
Symbols denote finite element simulation data (see \Appref{app:bucklingsim} for details). 
We set $\upsilon_\eff \Reqeq 0.3$ in the simulations. 
The solid curve corresponds to the analytical result \Eqref{eqn: the rescaled buckling pressure of orthotropic spheres} subtracting a constant offset of $0.0738$. 
} 
\label{fig: Buckling pressure of orthotropic spheres} 
\end{figure}

We compared our theoretical result against simulation results for the buckling load of orthotropic spherical shells with $\lambda > 1$, which were generated following the computational approach reported in \Refref{Herrmann2019} (see \Appref{app:bucklingsim} for details). To isolate the explicit dependence of the buckling pressure on the anisotropy parameter, theory and simulation values were rescaled by the classical buckling pressure of an isotropic sphere with the same radius and elastic parameters $\{E_1,\upsilon_{12}\}$: 
\begin{equation} 
  p_\text{sc}^\text{M} 
\coloneqq 
- \frac{ 2E_1 }{ 
  \sqrt{ 
  3\left( 1 - \upsilon_{12}^2 \right) 
  } 
  } 
  \left(\frac{t}{R}\right)^2. 
\label{eqn: the pressure scale by Munglani} 
\end{equation} 
Using this pressure scale, the rescaled prediction for the buckling pressure of spheres with $\lambda > 1$ is
\begin{equation} 
          \eta_\text{c, sph}^\text{M} 
\coloneqq \frac{ p_\text{c, sph} }{ p_\text{sc}^\text{M} } 
        = \frac{1}{ \sqrt{\lambda} }\sqrt{ 
          \frac{1}{1 - \upsilon_\eff^2}\left( 
          1 
        - \frac{\upsilon_\eff^2}{\lambda} 
          \right) 
          }, 
\label{eqn: the rescaled buckling pressure of orthotropic spheres} 
\end{equation} 
which is plotted as a solid line in \Figref{fig: Buckling pressure of orthotropic spheres}. 
We found that upon subtracting a constant offset of $0.0738$, the theoretical result successfully captures the dependence of the buckling pressure on the anisotropy parameter. 
The constant offset, which comes from matching the simulated buckling pressure for an isotropic sphere ($\lambda = 1$) and the corresponding known theoretical expression~\cite{Zoelly1915Ueber}, is well within the expected deviation between theory and simulations due to factors such as imperfection sensitivity. 
\Referenceref{Herrmann2019} also reported buckling pressures for orthotropic spherical shells with  $\lambda < 1$, for which buckling was observed to first occur near the two poles where the type of orthotropy is not rectilinear but polar. 
The rescaling transformation does not apply to this form of anisotropy, so we cannot predict the global buckling pressure in this parameter region using our approach. 
\paragraph{Internal Buckling Pressure.} \label{par: spheroids, buckling load, internal} 
\def\internal{ \mathrm{int} } 
In our previous work~\cite{Wenqian2021}, we demonstrated qualitatively that because of the sign switch of the prestress component $\sigma_0^{22}$ ($= \frac{1}{2}pR_2\left(1 + \beta_0\right)$) at $\beta_0 = -1$, it is possible for a highly oblate spheroidal shell with $\beta_0 < -1$ to buckle under a high enough internal pressure ($p >0$, $\p > 0$ in our convention) due to compressive stresses along its equator. 
Using \Eqref{eqn: pressurized k, spheroids, the closed form}, we are able to identify that pressure exactly. 
For $\beta_0 < -1$ ($\alpha < 0$) and $\p > 0$, the function $k(\p(\lambda), \beta_0)$ vanishes when $1 + \alpha\p = 0$, or equivalently, when $\p$ reaches the internal buckling pressure 
\refstepcounter{equation} \label{eqn: the internal buckling pressure} 
\begin{equation} 
\cp^\internal 
   \coloneqq 
     -\frac{1}{\alpha} 
   =  \frac{1 - \beta_0}{ \abs{1 + \beta_0} } 
   > 0. 
\tag{\theequation, a} 
\label{eqn: the internal buckling pressure, dimensionless} 
\end{equation} 
(This behavior arises from the property  $\displaystyle \lim_{x \to +\infty} \abs{ \sqrt{x}\, \EllipticF{\vartheta}{x} } = +\infty$.) 
Restoring the physical units gives 
\begin{equation} 
  \CP^\internal 
= \frac{ 4\sqrt{D' Y'} }{R_2^2 - 2R_1R_2} 
\tag{\theequation, b} 
\label{eqn: the internal buckling pressure, dimensionful} 
\end{equation} 
(cf.\ \Eqsref{eqn: the buckling pressure of orthotropic spheroids}). 
\Eqnsref{eqn: the internal buckling pressure} are consistent with predictions by Tovstik and Smirnov for the internal buckling pressure of highly oblate isotropic spheroidal shells~\cite{Tovstik_Internal_Buckling}, and yet again show that the orthotropic shell response is dictated by replacing the isotropic elastic constants with their geometric-mean counterparts $D'$ and $Y'$. \section{Discussion} \label{sec: discussions} 
We have established that under a particular coordinate transformation (\Eqref{eqn: the underlying coordinate transformation}), which we termed the rescaling transformation, an orthotropic shallow shell can be treated locally as an isotropic one of a different geometry. 
The principle underlying the rescaling transformation---mapping an anisotropic system to an isotropic one by rescaling the coordinate system used---has also been used in other contexts, e.g., the anisotropic $XY$-model~\cite{Schneider1992_Rescaling_XY}. 
The rescaling transformation enabled us to obtain analytical expressions for the local mechanical properties of orthotropic spheroidal and cylindrical shells, such as their buckling load (\Eqsref{eqn: the buckling pressure of orthotropic spheroids}, \eqref{eqn: critical axial stress of orthotropic cylinder} and \eqref{eqn: critical circumferential stress of orthotropic cylinder}) and indentation stiffness (\Eqsref{eqn: pressurized k, spheroids, the closed form} and \eqref{eqn: zero-pressure k, long cylinders, scaling form}), directly from using the corresponding isotropic results.

Besides its mathematical convenience that engendered new exact results for orthotropic shells, the transformation also helped to quantify the separate effects of geometry and material anisotropy on these local mechanical properties. 
We demonstrated that when the principal directions of curvature and material anisotropy are aligned, the two forms of anisotropy are effectively decoupled---our expressions factor into terms that capture the elasticity, multiplied with terms that incorporate the shell geometry. 
A consequence of this decoupling is that an orthotropic shell can have identical local mechanical properties as an isotropic shell with the same local geometry, and with appropriately chosen elastic parameters that render the material contributions identical as well.
This fact was previously recognized and exploited in the geometric-mean isotropic (GMI) approach to studying orthotropic cylinders, in which the orthotropic material was replaced with an isotropic material whose elastic constants are geometric means of the orthotropic values~\cite{Paschero2009Ortho_Cylinders}.
Our mapping rigorously establishes the equivalence of the two problems when the orthotropic in-plane shear modulus satisfies the Huber form ($G_\Huber$ given by \Eqref{eqn: the Huber form}).

The separation of geometry and material properties allowed us to use the results of our previous work, \Refref{Wenqian2021}, to calculate the geometric contribution to local mechanical properties.
The effects of material anisotropy, which we have derived in this work, differ depending on whether or not the deformation considered is localized. 
In the case of a localized deformation, the anisotropic elastic constants combine in the form of geometrical mean as in the GMI approach; the resulting combinations serve the role of effective isotropic elastic constants. 
However, if the deformation is not localized, like the case of indenting a cylinder at zero pressure (Section \ref{par: cylinders, stiffness, pressureless}), local mechanical properties can also depend on other dimensionless combinations of the anisotropic elastic constants, such as their ratio (see \Eqref{eqn: zero-pressure k, long cylinders, scaling form}). 
In such cases, the GMI approximation is no longer appropriate since the geometric means of the elastic constants do not capture all the material effects. 

We assumed throughout this paper that the orthotropic in-plane shear modulus $G_{12}$ is given by the Huber form ($G_\Huber$, \Eqref{eqn: the Huber form}).
While the Huber form is widely used and its validity has been verified for several different forms of orthotropic materials~\cite{Huber1923,Panc1975,Cheng1984Composite_Cylinders}, it is known that some properties of general orthotropic materials, such as tristability~\cite{Vidoli2008_Tristability}, require a departure from the Huber form.
Our work provides theoretical insight on why this is the case: any local elastic behavior of a Huber-form orthotropic shell can, through our mapping, also be observed in an isotropic shell of the same geometry.
Our mapping cannot be used to calculate properties of general orthotropic materials that would not be observed in an isotropic shell.
However, we envision that the rescaling transformation can be adapted to shells made of general orthotropic materials: In these cases, $G_{12}$ is in general a free parameter, and a torsion-like term with coupling constant proportional to $(G_{12} - G_\Huber)$ needs to be added into the governing equations in addition to the terms that are derived from isotropic shells. This additional term could be analyzed as the driver of phenomena that have no counterparts in isotropic shells.

Besides the potential for extension to general (non-Huber form) orthotropic materials, our work points to a few additional directions for future investigations. 
First, recall from \textbf{\ref{subsec: a rescaling transformation}} that the rescaling transformation only applies in the case where the axes of curvature and material anisotropy perfectly coincide (\Figref{fig: orthotropic structures}\ (c)). 
When the two sets of axes are misaligned, i.e., they are locally related by a rotation of some angle, we expect that effects of geometry and material anisotropy can couple together, unlike the case we studied in this paper, potentially leading to richer stiffness behaviors that could be useful for structural design. 
In the case of spheroids for which the extrinsic curvature tensor $\bm{\shape{}}$ and the prestress tensor $\bm{\sigma}_0$ share the same principal axes, the effect of the mismatch between the two sets of axes can be captured by a second torsion-like term\footnote{Note that this term, which will contain second-order derivatives, is different from the term related to the Huber form $G_\Huber$ that is proportional to fourth-order derivatives.} whose coupling constant will be given by off-diagonal components of $\bm{\sigma}_0$ if one uses as the coordinate axes the principal material axes. 
Furthermore, we note that at the poles of the spheroids that we considered, material orthotropy becomes curvilinear. 
The rescaling transformation does not apply in this case because of the complicated form that the biharmonic operator takes in polar coordinates~\cite{Anisotropy_Text}. 
It remains an open question what the buckling pressure and indentation stiffness of shells of curvilinear orthotropy are. 
Knowing these can shed some light on morphogenesis~\cite{Herrmann2019}, such as the reason why an apple develops a cusp~\cite{Chakrabarti2021_Apple}.
\begin{acknowledgments} 
The work of W.~S.\ was partly supported by the Lokey Doctoral Science Fellowship from the University of Oregon.
R.~V.\ acknowledges financial support for the development of the simulations of a pressurized orthotropic spherical shell by ETH Z\"{u}rich through ETHIIRA Grant no.\ ETH-03 10-3. 
W.~S.\ also would like to thank Abhijeet Melkani for discussions on evaluation of the stiffness integral.
\end{acknowledgments}

\onecolumngrid 
\appendix 

\newenvironment{remark}{\noindent{\textbf{\textit{Remark.}}}}{} 
\renewcommand{\theequation}{\thesection.\arabic{equation}} 
\newpage 
\section{Unrescaled Linearized Shallow-Shell Equations} \label{appendix: unrescaled equations} 
This appendix contains expressions for linearized (rectilinearly) anisotropic shallow-shell equations~\footnote{
The fact that the original nonlinear shallow-shell equations can be linearized implies that rectilinearly orthotropic shells can deform uniformly under a uniform pressure, at least in an approximate sense. 
} written in terms of unrescaled coordinates (which are unprimed in this paper) without using tensor notation. 
The dimensionless version of the equations has been derived in Ref.~\onlinecite{Nemeth1994SSEquations}; we are here going to restore physical units. 
\par 
For shells made of orthotropic materials, the compatibility equation takes the following form: 
\def\sqrlam   { \sqrt{\lambda} } 
\def\invsqrlam{ \frac{1}{\sqrlam} } 
\begin{equation} 
  \sqrlam\pdfrac[4]{ \Phi(x, y) }{x} 
+ 2E_\eff\left( 
  \frac{1}{ 2G_{12} } - \frac{ \upsilon_{12} }{E_2} 
  \right) 
  \frac{ \partial^4{ \Phi(x, y) } }{ 
  { \partial{x} }^2\, 
  { \partial{y} }^2 
  } 
+ \invsqrlam\pdfrac[4]{ \Phi(x, y) }{y} 
= Y'\left( 
  \frac{1}{R_2}\pdfrac[2]{ w(x, y) }{x} 
+ \frac{1}{R_1}\pdfrac[2]{ w(x, y) }{y} 
  \right), 
\end{equation} 
and the EOE is given by 
\begin{align} 
\begin{split} 
&   \sqrlam\pdfrac[4]{ w(x, y) }{x} 
  + 2\frac{1}{D'}\frac{t^3}{12}\left( 
    2G_{12} + \frac{ E_1\upsilon_{12} }{ 1 - \upsilon_{12}\upsilon_{21} } 
    \right) 
    \frac{ \partial^4{ w(x, y) } }{ 
    { \partial{x} }^2\, 
    { \partial{y} }^2 
    } 
  + \invsqrlam\pdfrac[4]{ w(x, y) }{y}\, 
  + \\[0.25 em] 
& \quad\quad 
  + \frac{1}{D'}\left( 
    \frac{1}{R_2}\pdfrac[2]{ \Phi(x, y) }{x} 
  + \frac{1}{R_1}\pdfrac[2]{ \Phi(x, y) }{y} 
    \right) 
  = \frac{1}{D'}\left( 
     \sigma^{11}_0t\pdfrac[2]{ w(x, y) }{x} 
  + 2\sigma^{12}_0t\mpdfrac  { w(x, y) }{x}{y} 
  +  \sigma^{22}_0t\pdfrac[2]{ w(x, y) }{y} 
    \right). 
\end{split} 
\end{align} 
Let $G_{12} \stackrel{!}{=} \frac{E_\eff}{ 2(1 + \upsilon_\eff) }$, i.e., assuming that the Huber form applies. 
We notice the following simplifications: 
\begin{equation} 
  \frac{1}{ 2G_{12} } - \frac{ \upsilon_{12} }{E_2} 
= \frac{1 + \upsilon_\eff}{E_\eff} 
- \frac{    \upsilon_\eff}{E_\eff} 
= \frac{1}                {E_\eff} 
\end{equation} 
and 
\begin{equation} 
  \frac{t^3}{12}\left( 
  2G_{12} + \frac{ E_1\upsilon_{12} }{ 1 - \upsilon_{12}\upsilon_{21} } 
  \right) 
= \frac{t^3}{12}\left( 
  \frac{E_\eff}             {1 + \upsilon_\eff} 
+ \frac{E_\eff\upsilon_\eff}{1 - \upsilon_\eff^2} 
  \right) 
= \frac{E_\eff t^3}{ 12\left(1 - \upsilon_\eff^2\right) } 
\equiv 
  D'. 
\end{equation} 
The two shallow-shell equations then reduce to 
\refstepcounter{equation} \label{eqn: the unrescaled EOEs} 
\begin{equation} 
  \left( 
            \sqrt[4]{\lambda}  \pdfrac[2]{}{x} 
+ \frac{1}{ \sqrt[4]{\lambda} }\pdfrac[2]{}{y} 
  \right)^2 
  { \Phi(x, y) } 
\eqqcolon 
  \opL{ \Phi(x, y) } 
= Y'\Vlasov{ w(x, y) } 
  \tag{\theequation, a} 
\end{equation} 
and 
\begin{equation} 
  D'\opL{ w(x, y) } 
+ \Vlasov{ \Phi(x, y) } 
=  \sigma^{11}_0t\pdfrac[2]{ w(x, y) }{x} 
+ 2\sigma^{12}_0t\mpdfrac  { w(x, y) }{x}{y} 
+  \sigma^{22}_0t\pdfrac[2]{ w(x, y) }{y}, 
  \tag{\theequation, b} 
\end{equation} 
where 
$
\Vlasov \equiv \frac{1}{R_2}\pdfrac[2]{}{x} 
             + \frac{1}{R_1}\pdfrac[2]{}{y} 
$
denotes the Vlasov operator. 
Combining the two equations, we obtain 
\begin{equation} 
  D'\opL^2   { w(x, y) } 
+ Y'\Vlasov^2{ w(x, y) } 
=   \opL  \left( 
   \sigma^{11}_0t\pdfrac[2]{ w(x, y) }{x} 
+ 2\sigma^{12}_0t\mpdfrac  { w(x, y) }{x}{y} 
+  \sigma^{22}_0t\pdfrac[2]{ w(x, y) }{y} 
  \right). 
\label{eqn: the unrescaled eqn for w} 
\end{equation} \newpage 
\section{Mechanical Properties at the Poles of an Orthotropic Spheroid.} \label{appendix: pole calculations} 
In this appendix, we will derive the indentation stiffness at the poles of an orthotropic spheroid in the absence of pressure. 
Recall that the material orthotropy pattern is curvilinear at the poles. 
The result is obtained in two ways, first by a qualitative energy-balance argument which is then supported by analytically solving the governing linearized equations of equilibrium (EOEs). 
We finish the appendix with a short discussion about what will happen if the spheroid is pressurized. 
\subsection{Zero-Pressure Indentation Stiffness.} 
\subsubsection{The Energy-Balance Argument.} 
Landau and Lifshitz first used this approach to obtain the indentation stiffness and buckling pressure of an isotropic spherical shell~\cite{LL_Elasticity}. 
We here modify their approach to include polar material orthotropy. 
\begin{figure}[H] 
\centering 
\includegraphics[width = 0.36\textwidth]{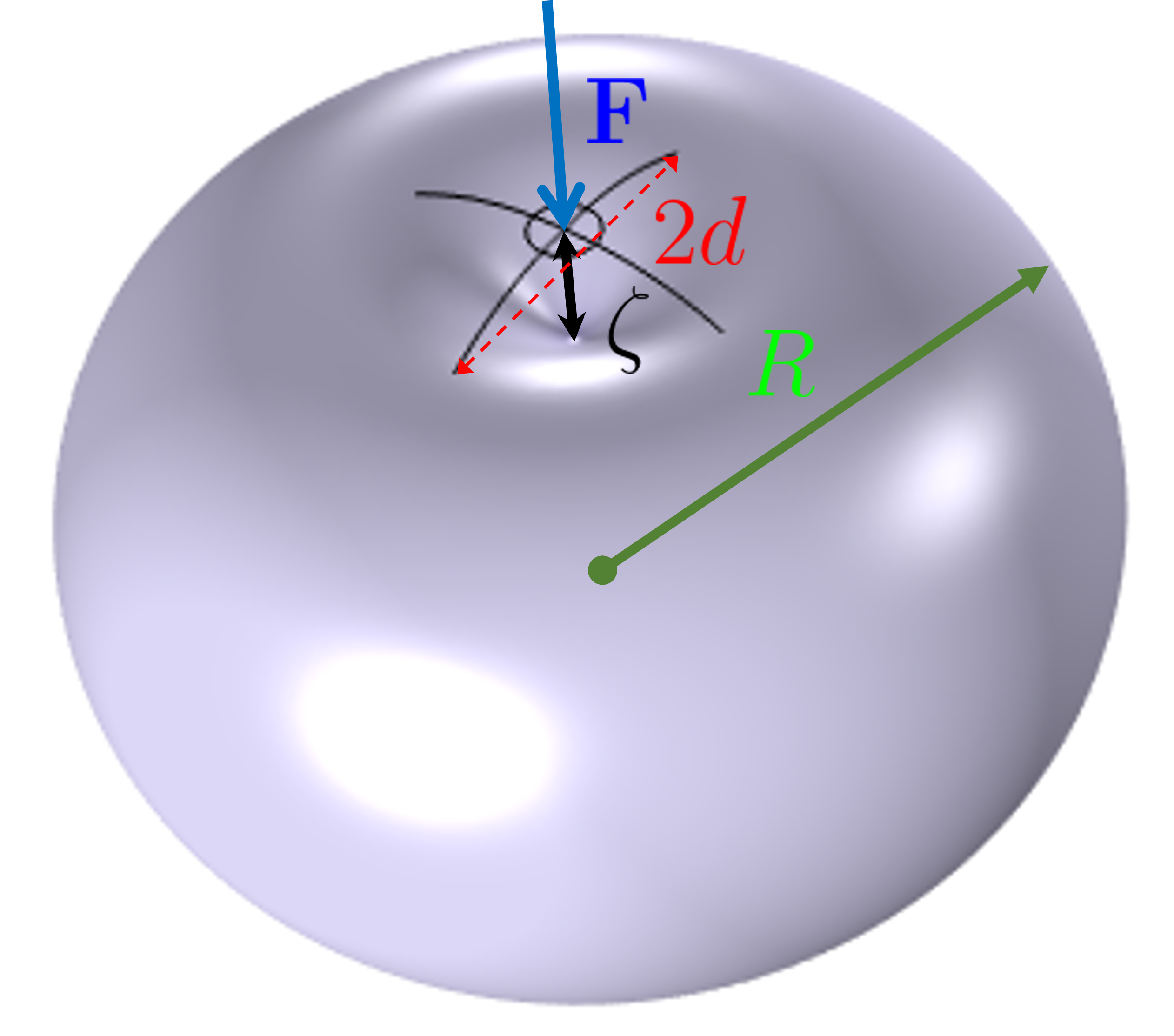} 
\caption{
Indenting a spherical shell of radius $R$ near one of its poles. 
A point load, $\mathbf F$, is applied right at the pole. 
The radius of the resulting deformed region is roughly $d$, and the vertical deflection along $\mathbf F$ is denoted by $\zeta$. 
} 
\label{fig appendix: Indentation schematic} 
\end{figure} 
\noindent \Figureref{fig appendix: Indentation schematic} depicts that a point load $\mathbf F$ is applied at one of a spheroid's poles, the center of a locally spherical region with radius $R$. 
The area of the resulting deformed region is of the order $d^2$ ($\sim d^2$). 
The deflection $\zeta$ varies significantly over a distance of $d$, which implies that the bending energy is $\sim E_r t^3\left(\frac{\zeta}{d^2}\right)^2d^2$, where $E_r$ denotes Young's modulus along the meridional direction. 
The reason why $E_r$ was used to estimate the bending energy is that from the cross-sectional view, \Figref{fig appendix: Indentation schematic}, shell bending mainly occurs in the meridional direction, while stretching happens in the zonal direction. 
\par 
Strain does not depend on $d$ and is $\sim \frac{\zeta}{R}$. 
The stretching energy is thus $\sim E_\theta t\left(\frac{\zeta}{R}\right)^2d^2$, and the total elastic energy is roughly 
\begin{equation} 
U \sim \frac{E_r t^3\zeta^2}{d^2} + \frac{E_\theta t\zeta^2}{R^2}d^2. 
\label{eqn: E_tot in LL argument} 
\end{equation} 
The global minimum of $U$ can be rapidly obtained by recalling the AM-GM inequality: 
\begin{equation} 
     U_{\min} 
\sim \frac{2\sqrt{E_r E_\theta}t^2\zeta^2}{R} 
   = 2\sqrt{ 
     \left(\frac{E_r      t^3\zeta^2}{d^2}   \right) 
     \left(\frac{E_\theta t  \zeta^2}{R^2}d^2\right) 
     } 
\leq \frac{E_r      t^3\zeta^2}{d^2} 
   + \frac{E_\theta t  \zeta^2}{R^2}d^2 
\sim U. 
\label{eqn: using the AM-GM inequality} 
\end{equation} 
Varying $U_{\min}$ with respect to $\zeta$ and equating the result to $F\, \delta{\zeta}$, the variation of the work done by the point load, we find the deflection $\zeta \sim \frac{FR}{4\sqrt{E_r E_\theta}t^2}$ and hence the indentation stiffness 
\begin{equation} 
     k_{p = 0}^\text{pole} 
   = \frac{F}{\zeta} 
\sim \frac{4\sqrt{E_r E_\theta}t^2}{R}, 
\label{eqn: indentation stiffness, pole, zero pressure} 
\end{equation} 
which agrees with \Eqref{eqn: zero-pressure stiffness} up to a factor of two. 
As this argument explicitly shows, although the local symmetry at the equator (see \textbf{\ref{subsec: the local indentation stiffness}}) completely breaks down at the poles, i.e., the two orthogonal directions now become curvilinear and hence distinguishable, the geometric-mean dependence persists and stems from balancing the bending and stretching energies. 
\subsubsection{The Analytical Approach.} 
\Eqnref{eqn: indentation stiffness, pole, zero pressure} can also be obtained by solving the EOEs that govern the deformations of a curvilinearly orthotropic shallow spherical shell. 
The full nonlinear EOEs can be found in, for example, \Refref{Anisotropy_Text}. 
Since we only consider small deformations due to a point load at the center of the shell, it is reasonable to linearize these equations and further assume, from a symmetry point of view, that the deformations of interest are axisymmetric, i.e., do not vary along the azimuthal direction. 
In this case, the governing equations reduce to~\cite{Indian_Ortho_Sph_Cap} 
\refstepcounter{equation} \label{eqn: linearized equations of SST, poles} 
\def\invsqrlam{ \frac{1}{ \sqrt{\lambda} } } 
\begin{align*} 
& D_r               \Bessel{\invsqrlam}{    y(r) } + \frac{    y(r) }{R} = -\frac{F}{2\pi}\frac{1}{r} \tag{\theequation, a} \\[0.25 em] 
& \frac{1}{Y_\theta}\Bessel{\invsqrlam}{ \phi(r) } - \frac{ \phi(r) }{R} = 0,                         \tag{\theequation, b} 
\end{align*} 
where 
$
          D_r 
\coloneqq \frac{E_r t^3}{ 12\left(1 
        - \upsilon_{r \theta} 
          \upsilon_{\theta r} 
          \right) } 
$ is the bending stiffness along the meridional direction; $Y_\theta \coloneqq E_\theta t$ the Young's modulus in the zonal direction; and $\lambda \coloneqq \frac{E_r}{E_\theta}$ the anisotropy parameter in this case. 
That $D_r$ and $Y_\theta$ show up in the governing equations supports our previous observation that shell bending and stretching occur in different directions. 
The fields $y$ and $\phi$ are the first derivative of the normal displacement $u_3$ and the Airy stress function $\Phi$, respectively: 
$y    \coloneqq \odfrac{u_3 }{r}$, and 
$\phi \coloneqq \odfrac{\Phi}{r}$, where $r$ is the distance away from the pole. 
The operator 
$
       \Bessel{\nu}{} 
\equiv            \odfrac[2]{}{r} 
     + \frac{1}{r}\odfrac[ ]{}{r} 
     - \left(\frac{\nu}{r}\right)^2 
     \mbox{ } (\nu \in \mathbb C) 
$ is the Bessel differential operator. 
It is known that Bessel functions of the first kind with order $\nu$ (denoted by $J_\nu$) are its eigenfunctions. 
This motivates us to solve \Eqsref{eqn: linearized equations of SST, poles} using the Hankel transform. 
\paragraph{Hankel Transform.} Roughly speaking, Hankel transform is like Fourier transform in polar coordinates and is often used to solve \emph{linear axisymmetric} differential equations. 
The Hankel transform of a well-behaved axisymmetric function $f(r)$ is given by~\cite{Piessens_Hankel} 
\begin{equation} 
       \hat{f}_\nu(k) 
\equiv \mathscr{H}_\nu\left\{f(r)\right\}(k) 
     = \int_0^{+\infty} r\, \df{r}\, 
       f(r)J_\nu(kr). 
\end{equation} 
The inverse transform is given by 
\begin{equation} 
  f(r) 
= \int_0^{+\infty} k\, \df{k}\, 
  \hat{f}_\nu(k)J_\nu(kr). 
\end{equation} 
\par 
The Hankel transform of the Bessel operator, $\Bessel{\nu}{}$, is simply $-k^2$, which is independent of $\nu$. 
This can be most easily seen by recalling the definition of the Bessel differential equation: 
\begin{equation} 
\left(\Bessel{\nu}{} + k^2\right)J_\nu(kr) = 0. 
\end{equation} 
It follows that for an axisymmetric function $f(r)$, 
\begin{equation} 
  \Bessel{\nu}{ f(r) } 
= \Bessel{\nu}{ 
  \int_0^{+\infty} k\, \df{k}\, 
  \hat{f}_\nu(k)J_\nu(kr) 
  } 
= \int_0^{+\infty} k\, \df{k}\, 
  \left(-k^2\hat{f}_\nu(k)\right)J_\nu(kr). 
\end{equation} 
It is also straightforward to obtain the Hankel transform of the function $\frac{1}{r}$: By definition, 
\begin{equation} 
  \mathscr{H}_\nu\left\{\frac{1}{r}\right\}(k) 
= \int_0^{+\infty} r\, \df{r}\, 
  \frac{1}{r}J_\nu(kr) 
= \int_0^{+\infty} \df{r}\, J_\nu(kr) 
= \frac{1}{k}\int_0^{+\infty} \df{u}\, J_\nu(u) 
= \frac{1}{k}, 
\end{equation} 
where we have used the fact that for all $\nu$, 
\begin{equation} 
\int_0^{+\infty} \df{x}\, J_\nu(x) = 1. 
\end{equation} 
The Hankel transform of \Eqsref{eqn: linearized equations of SST, poles} is hence 
\refstepcounter{equation} \label{eqn: the HTed EOEs} 
\begin{align*} 
  - D_r        k^2\hat{y   }_\invsqrlam(k) 
  + \frac{1}{R}   \hat{\phi}_\invsqrlam(k) 
& = -\frac{F}{2\pi}\frac{1}{k} 
    \tag{\theequation, a} 
    \label{eqn: the HTed EOE, 1} 
    \\[0.25 em] 
  - \frac{1}{Y_\theta}k^2\hat{\phi}_\invsqrlam(k) 
  - \frac{1}{R}          \hat{y   }_\invsqrlam(k) 
& = 0. 
    \tag{\theequation, b} 
    \label{eqn: the HTed EOE, 2} 
\end{align*} 
Substituting \Eqref{eqn: the HTed EOE, 2} into \Eqref{eqn: the HTed EOE, 1} to eliminate $\hat{\phi}_\invsqrlam(k)$, we get, after applying the inverse transform, 
\begin{equation} 
  \odfrac{w}{r}(r) 
\eqqcolon 
  y(r) 
= \frac{F}{2\pi}\int_0^{+\infty} \df{k}\, 
  \frac 
  {k^2} 
  { \displaystyle D_r k^4 + \frac{Y_\theta}{R^2} } 
  J_\invsqrlam(kr). 
\label{eqn: the expression for y} 
\end{equation} 
To proceed, we impose the boundary conditions $w(0) = -\zeta$ and $\displaystyle \lim_{r \to +\infty} w(r) = 0$ which together give 
\begin{equation} 
  \int_0^{+\infty} \df{r}\, \odfrac{w}{r}(r) 
= \lim_{r \to +\infty} w(r) - w(0) 
= \zeta. 
\label{eqn: boundary conditions for w} 
\end{equation} 
Combining \Eqsref{eqn: the expression for y} and \eqref{eqn: boundary conditions for w}, we finally attain the following relation between $\zeta$ and $F$: 
\begin{align} 
\begin{split} 
    \zeta 
& = \int_0^{+\infty} \df{r}\, 
    \frac{F}{2\pi}\int_0^{+\infty} \df{k}\, 
    \frac 
    {k^2} 
    { \displaystyle D_r k^4 + \frac{Y_\theta}{R^2} } 
    J_\invsqrlam(kr) 
  = \frac{F}{2\pi}\int_0^{+\infty} \df{k}\, 
    \frac 
    {k^2} 
    { \displaystyle D_r k^4 + \frac{Y_\theta}{R^2} } 
    \int_0^{+\infty} \df{r}\, J_\invsqrlam(kr) 
    \\[0.25 em] 
& = \frac{F}{2\pi}\int_0^{+\infty} \df{k}\, 
    \frac 
    {k} 
    { \displaystyle D_r k^4 + \frac{Y_\theta}{R^2} } 
  = F\frac{R}{ 8\sqrt{D_r Y_\theta} }. 
\end{split} 
\label{eqn: the expression for zeta} 
\end{align} 
From \Eqref{eqn: the expression for zeta}, we can get the indentation stiffness: 
\begin{equation} 
  k 
\coloneqq 
  \frac{F}{\zeta} 
= \frac{ 8\sqrt{D_r Y_\theta} }{R} 
= \frac{ 4\sqrt{E_r E_\theta}t^2} 
  { 
  \sqrt{ 
  3\left( 
  1 
- \upsilon_{r \theta} 
  \upsilon_{\theta r} 
  \right) 
  } 
  } 
  \frac{1}{R} 
\label{eqn: zero-pressure stiffness, sphere, poles} 
\end{equation} 
(cf. \Eqref{eqn: zero-pressure stiffness}). 
This proves our claim in the main text (see \textbf{\ref{subsec: the local indentation stiffness}}). 
As the figure below shows, our analytical expression agrees well with numerical simulations using \COMSOL. 
\begin{figure}[H] 
\centering 
\includegraphics[width = 0.45\textwidth]{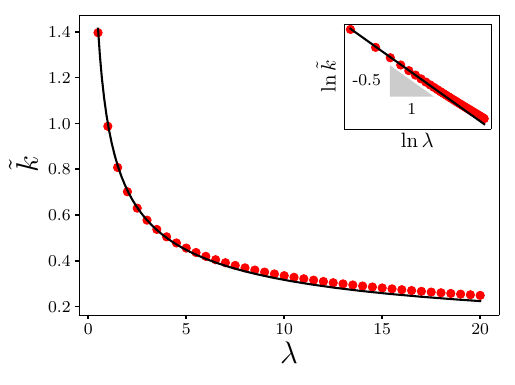} 
\caption{
Zero-pressure indentation stiffness of an orthotropic sphere at its poles as a function of the anisotropy parameter $\lambda$. 
Symbols denote data obtained from \COMSOL{} simulations. 
Solid curves correspond to the analytical expression \Eqref{eqn: zero-pressure stiffness, sphere, poles}. 
Indentation stiffness is scaled by 
$
\frac{4E_r t^2}{ \sqrt{ 3\left(1 - \upsilon_{r \theta}\upsilon_{\theta r}\right) } }\frac{1}{R} 
$. 
The inset shows the same data on double-log scale. 
} 
\label{fig appendix: zero-pressure stiffness for curvilinearly orthotropic spheres} 
\end{figure} 
\subsection{Pressurized Orthotropic Spheroids.} 
Recall the fact that near its poles, a spheroid is locally spherical. 
Therefore, the following discussions are centered around curvilinearly orthotropic spherical shells. 
\par 
Unlike its isotropic counterpart, a curvilinearly orthotropic sphere does not deform uniformly under a constant pressure. 
This can be seen by substituting the membrane solution 
\begin{equation} 
\begin{cases} 
y   _\text{m}(r) = 0, 
                   \\[0.25 em] 
\phi_\text{m}(r) = \displaystyle 
                   \frac{1}{2}pR r, 
\end{cases} 
\end{equation} 
into the nonlinear shallow-shell equations~\cite{Indian_Ortho_Sph_Cap} 
\refstepcounter{equation} \label{eqn: equations of SST, poles} 
\begin{align*} 
    D_r 
    \Bessel{\invsqrlam}{ y(r) } 
  - \frac{ \phi(r) }{r}\left(y(r) - \frac{r}{R}\right) 
& = \frac{1}{2}p r 
    \tag{\theequation, a} 
    \label{eqn: the EOE, poles} 
    \\[0.25 em] 
    \frac{1}{Y_\theta} 
    \Bessel{\invsqrlam}{ \phi(r) } 
  + \frac{1}{2}\frac{ y(r) }{r}\left(y(r) - \frac{2r}{R}\right) 
& = 0. 
    \tag{\theequation, b} 
    \label{eqn: the compatibility equation, poles} 
\end{align*} 
\Eqnref{eqn: the compatibility equation, poles} gives $\left(1 - \frac{1}{\lambda}\right)\frac{pR}{Y_\theta} = 0$ which only holds in the isotropic case ($\lambda = 1$). 
Moreover, we notice that $\left(1 - \frac{1}{\lambda}\right)\frac{pR}{Y_\theta}$ switches its sign at $\lambda = 1$. 
The presence of this term illustrates the fact that upon being pressurized, spheres with a curvilinear orthotropy pattern deform differently depending on whether $E_r > E_\theta$ or the other way around~\cite{Reissner1958, Sobota2019_Ortho_Shapes}. 
Therefore, the term cannot be ignored in general, and linearization using the membrane solution thus generally fails for these shells. 
\par 
In fact, as Reissner has demonstrated, for pressurized curvilinearly orthotropic spheres, both the displacement field $w(r)$ and the Airy stress function $\Phi(r)$ scale as $r^{\invsqrlam + 1}$ near the origin~\cite{Reissner1958}. 
As a result, the actual stress, $\norm{ \bm{\sigma}(r) }t \sim \frac{ \Phi(r) }{r^2}$ will have the power-law behavior $r^{\invsqrlam - 1}$; that is, depending on the magnitude of $\lambda$, the stress at the poles will either vanish ($\lambda < 1$) or explode ($\lambda > 1$). 
This stress singularity makes it challenging to derive the indentation stiffness and buckling pressure of pressurized curvilinearly orthotropic spheres in general. 
\par
\begin{figure}[H] 
\centering 
\includegraphics[width = 0.45\textwidth]{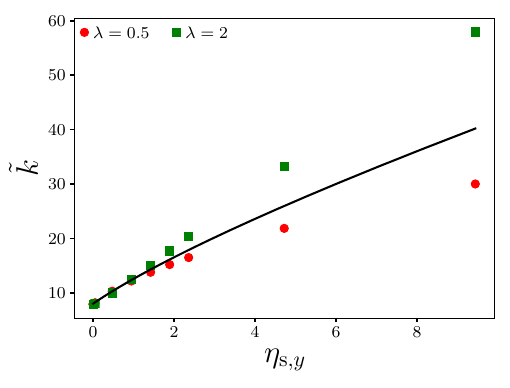} 
\caption{
Indentation stiffness of two orthotropic spheres with different degrees of anisotropy as a function of pressure. 
Symbols denote data obtained from \COMSOL{} simulations. 
The solid curve corresponds to the analytical expression \Eqref{eqn: stiffness at the poles of a sphere}. 
Indentation stiffness is scaled by 
$
\frac{ \sqrt{D_r Y_\theta} } {R} 
$. 
} 
\label{fig appendix: stiffness of curvilinearly orthotropic spheres} 
\end{figure}

 
However, for sufficiently low pressures, such that the approximation $\left(1 - \frac{1}{\lambda}\right)\frac{pR}{Y_\theta} \approx 0$ can be safely made, following the same procedure as in the pressureless case, we obtain 
\begin{align} 
\begin{split} 
\zeta & = \frac{F}{2\pi}\int_0^{+\infty} \df{k}\, 
          \frac 
          {k} 
          { \displaystyle D_r k^4 + \frac{pR}{2}k^2 + \frac{Y_\theta}{R^2} } 
        = \frac{F}{4\pi}\sqrt{ \frac{R^2}{D_r Y_\theta} } 
          \int_0^{+\infty} 
          \frac{ \df{u} }{u^2 + 2\p u + 1} 
          \\[0.25 em] 
      & = F\frac{R}{ 8\sqrt{D_r Y_\theta} } 
          \frac 
          { \displaystyle 1 - \frac{2}{\pi}\arcsin{\p} } 
          { \sqrt{1 - \p^2} }, 
\end{split} 
\label{eqn: the expression for zeta, the pressurized case} 
\end{align} 
where $\p \coloneqq \frac{pR^2}{ 4\sqrt{D_r Y_\theta} }$. 
The indentation stiffness in this case is hence 
\begin{equation} 
       k 
\equiv \frac{F}{\zeta} 
     = \frac{ 8\sqrt{D_r Y_\theta} } {R} 
       \frac 
       { \sqrt{1 - \p^2} } 
       { \displaystyle 1 - \frac{2}{\pi}\arcsin{\p} }. 
\label{eqn: stiffness at the poles of a sphere} 
\end{equation} 
Note that \Eqref{eqn: stiffness at the poles of a sphere} is still invariant under interchange of labels $r$ and $\theta$. 
Mathematically, this means that $k(\lambda) = k\left(\frac{1}{\lambda}\right)$. 
This analytical insight is confirmed by \COMSOL{} simulations, as \Figref{fig appendix: stiffness of curvilinearly orthotropic spheres} shows. 
At low pressures, such that $\frac{pR}{Y_\theta} \approx 0$, indentation stiffness of the two orthotropic spheres is basically identical to each other; however, when the scaled pressure increases to order one ($\p \sim 1$), we start to see deviations from the theory. 
The fact that the two sets of data fall onto different sides of the theory curve is a result of the term $\left(1 - \frac{1}{\lambda}\right)\frac{pR}{Y_\theta}$ being non-negligible. 
 \newpage 
\section{A Derivation for the Zero-Pressure Indentation Stiffness of Long Cylinders} \label{appendix: Yuan's analysis} 
In this appendix, we will combine Yuan's approach~\cite{Yuan1946_Cylinders} and the rescaling transformation to solve the linearized shallow-shell equation for long cylindrical shells: 
\refstepcounter{equation} \label{eqn: the SS eqn for cyls} 
\begin{equation} 
  D'\opL^2{ w(x, s) } 
+ \frac{Y'}{R^2}\pdfrac[4]{ w(x, s) }{x} 
=   \opL  { q(x, s) } 
\tag{\theequation, a} 
\label{eqn: the SS eqn for cyls, original} 
\end{equation} 
(see Eq.~\eqref{eqn: the unrescaled eqn for w}) or equivalently, 
\begin{equation} 
  D'\opL'^2{ w'(x', s') } 
+ \frac{Y'}{R'^2}\pdfrac[4]{ w'(x', s') }{x'} 
=   \opL'  { q'(x', s') }. 
\tag{\theequation, b} 
\label{eqn: the SS eqn for cyls, rescaled} 
\end{equation} 
\subsection{Yuan's Approach} 
In short, the approach by Yuan has two main distinctive features compared with our analysis in Ref.~\onlinecite{Wenqian2021}. 
First, along the circumferential direction (associated with the coordinate $s$), a Fourier series defined on $(-\pi R, \pi R]$, instead of a Fourier transform, was used: More specifically, a well-behaved function $f(x, s)$ can be written as 
\begin{equation} 
  f(x, s) 
= \int_{-\infty}^{+\infty} \frac{ \df{k} }{2\pi}\, 
  \hat{f}(k, s)e^{\imunit kx} 
= \int_{-\infty}^{+\infty} \frac{ \df{k} }{2\pi}\, 
  \sum_{n = -\infty}^\infty 
  \hat{f}_n(k) 
  e^{ \imunit n\frac{s}{R} } 
  e^{ \imunit kx}. 
\label{eqn: FT + FS} 
\end{equation} 
Furthermore, if the function $f(x, s)$ is even in both $x$ and $s$, the above expression reduces to 
\begin{equation} 
  f(x, s) 
= 2\int_0^{+\infty} \frac{ \df{k} }{2\pi}\, \left[ 
  \frac{1}{2}\hat{f}_0(k) 
+ \sum_{n = 1}^\infty 
  \hat{f}_n(k) 
  \cos\left(n\frac{s}{R}\right) 
  \right] 
  \cos{kx}, 
\label{eqn: FT + FS, even func} 
\end{equation} 
where we have implicitly used the fact that the Fourier transform of an even function is even. 
For a separable function, i.e., $f(x, s) = X(x)S(s)$, Eq.~\eqref{eqn: FT + FS, even func} becomes 
\begin{equation} 
  f(x, s) 
= 2\left[ 
  \frac{1}{2}S_0 
+ \sum_{n = 1}^\infty 
  S_n\cos\left(n\frac{s}{R}\right) 
  \right] 
  \int_0^{+\infty} \frac{ \df{k} }{2\pi}\, 
  \hat{X}(k)\cos{kx}. 
\label{eqn: FT + FS, even and separable func} 
\end{equation} 
Second, Yuan did not use the Dirac delta function to model a concentrated load; instead, he first considered a uniformly distributed load over a rectangular region and then shrank the size of the region. 
\subsection{The Rescaling Transformation} 
We can thus write 
\begin{align} 
\begin{split} 
    w(x, s) 
& = \int_0^{+\infty} \frac{ \df{k} }{2\pi}\, \left[ 
     \hat{w}_0(k) 
  + \sum_{n = 1}^\infty 
    2\hat{w}_n(k) 
    \cos\left(n\frac{s}{R}\right) 
    \right] 
    \cos{kx} 
    \\[0.25 em] 
& = \sum_{n = 0}^\infty \left[ 
    ( 2 - \delta_{0n} ) 
    \int_0^{+\infty} \frac{ \df{k} }{2\pi}\, 
    \hat{w}_n(k) 
    \cos{kx} 
    \cos\left(n\frac{s}{R}\right) 
    \right] 
\end{split} 
\label{eqn: FT + FS of w} 
\end{align} 
since the normal displacement field $w(x, s)$ must be an even function in both $x$ and $s$ from a symmetry point of view. 
The constant load is applied on a rectangular region that is symmetric with respect to the origin; therefore, $q(x, s)$ is even and separable: $q(x, s) = X(x)Q(s)$, and 
\begin{equation} 
  q(x, s) 
= \sum_{n = 0}^\infty \left[ 
  ( 2 - \delta_{0n} ) 
  \int_0^{+\infty} \frac{ \df{k} }{2\pi}\, 
  \hat{X}(k)Q_n 
  \cos{kx} 
  \cos\left(n\frac{s}{R}\right) 
  \right]. 
\label{eqn: FT + FS of q} 
\end{equation} 
Letting the region be $R = \left\{ (x, s) \in [-\epsilon, \epsilon] \times [-c, c] \right\}$, we can then determine $\hat{X}(k)$ and $Q_n$. 
By definition, 
\begin{equation} 
  \hat{X}(k) 
= \int_{-\infty}^{+\infty} \df{x}\, 
  X(x)e^{-\imunit kx} 
= 2\int_0^\epsilon \df{x}\, \cos{kx} 
= 2\epsilon\sinc{k\epsilon}, 
\end{equation} 
and 
\begin{equation} 
  Q_n 
= \frac{2}{\pi R}\int_0^{\pi R} \df{s}\, Q(s)\cos\left(n\frac{s}{R}\right) 
= \frac{2}{\pi R}\int_0^c       \df{s}\, q_0 \cos\left(n\frac{s}{R}\right) 
= \frac{2}{\pi}\frac{c}{R}\sinc\left(n\frac{c}{R}\right)q_0. 
\end{equation} 
The intensity of the load is denoted by $q_0$, and the total force is hence $F = q_0A = 4q_0c\epsilon$. 
In the limits of $k\epsilon \to 0$ and $n\frac{c}{R} \to 0$, 
\begin{equation} 
\hat{X}(k) \approx 2\epsilon, 
                   \textAnd 
Q_n        \approx \frac{2}{\pi}\frac{c}{R}q_0. 
\end{equation} 
\bigskip\par 
\begin{remark} 
In Yuan's original formulation of the problem, there is an additional concentrated load being applied at the bottom of the cylinder ($s = \pm\pi R$). 
As a consequence, when computing the Fourier coefficients $Q_n$ for the original system, an extra term, 
\begin{equation} 
  \frac{2}{\pi R}\int_{\pi R - c}^{\pi R} \df{s}\, Q(s)\cos\left(n\frac{s}{R}\right) 
= (-1)^n Q_n, 
\end{equation} 
needs to be added. 
This leads to vanishing of the odd terms in the Fourier series. 
\end{remark} 
\bigskip\par 
Therefore, for a point load, 
\begin{equation} 
  q(x, s) 
\approx 
  \frac{F}{\pi R}\sum_{n = 0}^\infty \left[ 
  ( 2 - \delta_{0n} ) 
  \int_0^{+\infty} \frac{ \df{k} }{2\pi}\, 
  \cos{kx} 
  \cos\left(n\frac{s}{R}\right) 
  \right]. 
\label{eqn: FT + FS of q, a point load} 
\end{equation} 
Substituting Eqs.~\eqref{eqn: FT + FS of w} and \eqref{eqn: FT + FS of q, a point load} into Eq.~\eqref{eqn: the SS eqn for cyls, original}, we obtain, after some algebra, 
\begin{equation} 
  \hat{w}_n(k) 
= \frac{F}{\pi R}\frac 
  {\displaystyle 
  \left( 
            \sqrt[4]{\lambda}  k^2 
+ \frac{1}{ \sqrt[4]{\lambda} }\frac{n^2}{R^2} 
  \right)^2 
  } 
  {\displaystyle 
  D'\left( 
            \sqrt[4]{\lambda}  k^2 
+ \frac{1}{ \sqrt[4]{\lambda} }\frac{n^2}{R^2} 
  \right)^4 
+ \frac{Y'}{R^2}k^4 
  }. 
\label{eqn: Fourier coefficients for w} 
\end{equation} 
We now apply the rescaling transformation in Fourier space: $R \mapsto R' = \sqrt[4]{\lambda}R$, $k \mapsto k' = \sqrt[8]{\lambda}k$ and $n \mapsto n' = \sqrt[8]{\lambda}n$; Eq.~\eqref{eqn: Fourier coefficients for w} then reduces to 
\def\tpk{\tilde{k}'} 
\begin{equation} 
  \hat{w}_n(k) 
= \sqrt[4]{\lambda}\frac{1}{\pi}\frac{F R'^3}{D'}\frac 
  {\left(\tpk^2 + n'^2\right)^2} 
  {\left(\tpk^2 + n'^2\right)^4 + \gamma'\tpk^4}, 
\label{eqn: Fourier coefficients for w, rescaled} 
\end{equation} 
where $\tpk \coloneqq R' k' = \sqrt[8]{\lambda^3}R k$, which is dimensionless, and $\gamma' \coloneqq \frac{Y' R'^2}{D'}$ is the \FvK{} number for the rescaled system. 
We note that Eq.~\eqref{eqn: Fourier coefficients for w, rescaled} can also be attained by directly substituting into Eq.~\eqref{eqn: the SS eqn for cyls, rescaled} Fourier series and transforms that are written in terms of the rescaled variables, e.g., 
\begin{equation} 
  f(x', s') 
= \frac{1}{ \sqrt[8]{\lambda} }\int_{-\infty}^{+\infty} \frac{ \df{k'} }{2\pi}\, 
  \sum_{n = -\infty}^\infty 
  \hat{f}_{n'}(k') 
  e^{ \imunit n'\frac{s'}{R'} } 
  e^{ \imunit k' x'}. 
\label{eqn: FT + FS, rescaled} 
\end{equation} 
From Eq.~\eqref{eqn: Fourier coefficients for w, rescaled}, we can get the following expression for the inverse of the indentation stiffness: 
\begin{align} 
\begin{split} 
    \frac{1}{ k_\cyl^0(\lambda) } 
  \coloneqq 
    \frac{ w(0, 0) }{F} 
& = \frac{1}{2\pi^2}\sqrt[8]{\lambda}\frac{R'^2}{D'}\sum_{n = 0}^\infty \left[ 
    ( 2 - \delta_{0n} )\int_0^{+\infty} \df{u}\, 
    \frac 
    {\left(u^2 + n'^2\right)^2} 
    {\left(u^2 + n'^2\right)^4 + \gamma'u^4} 
    \right] 
    \\[0.25 em] 
& = \frac{1}{2\pi^2}\sqrt[8]{\lambda}\frac{R'^2}{D'} 
    \int_0^{+\infty} \frac{ \df{u} }{u^4 + \gamma'} 
  + \frac{1}{ \pi^2}\sqrt[8]{\lambda}\frac{R'^2}{D'}\sum_{n = 1}^\infty 
    \int_0^{+\infty} \df{u}\, 
    \frac 
    {\left(u^2 + n'^2\right)^2} 
    {\left(u^2 + n'^2\right)^4 + \gamma'u^4}. 
\end{split} 
\label{eqn: inverse stiffness, cyl} 
\end{align} 
\bigskip\par 
\begin{remark} 
Notice that the $n = 0$ mode does not lead to a divergence, unlike the situation in Ref.~\onlinecite{Wenqian2021} where the stiffness was written in terms of the following double integral: 
\begin{equation} 
  \frac{1}{ k_\cyl^0(\lambda = 1) } 
= \frac{1}{2\pi^2}\frac{R}{ \sqrt{D Y} } 
  \int_0^\frac{\pi}{2} \df{\theta}\, 
  \int_0^{+\infty} \frac{ \df{u} }{ u^2 + \cos^4{\theta} }, 
\end{equation} 
which diverges in the infrared limit ($\mathbf{u \to 0}$). 
As Yuan found, the contribution of the $n = 0$ mode to the indentation stiffness is in fact negligible compared to other modes; as a result, the first term on the right-hand side of Eq.~\eqref{eqn: inverse stiffness, cyl} can be neglected. 
\end{remark} 
\bigskip\par 
Equation~\eqref{eqn: inverse stiffness, cyl} takes the same form as Eq.~(10) in Ref.~\onlinecite{Yuan1946_Cylinders}, except for an extra factor of $\frac{1}{2}\sqrt[8]{\lambda}$. 
We can hence directly apply Yuan's final result, Eq.~(17), without actually evaluating the definite integrals in Eq.~\eqref{eqn: inverse stiffness, cyl}: 
\begin{equation} 
  \frac{1}{ k_\cyl^0(\lambda) } 
\approx 
  \frac{1}{2\pi}\sqrt[8]{\lambda} 
  \frac{ 3\sqrt{2}\left(1 - \upsilon_\eff^2\right) }{E_\eff}\frac{R'^2}{t^3}\sum_{n = 1}^\infty 
  \frac{1}{n'^3}\frac{ \sqrt{1 + \Xi_n} }{\Xi_n}, 
\end{equation} 
where 
$
\Xi_n^2 \coloneqq 1 + \frac{ 3\left(1 - \upsilon_\eff^2\right) }{4n'^4}\left(\frac{R'}{t}\right)^2 
                = 1 + \frac{ 3\left(1 - \upsilon_\eff^2\right) }{4n ^4}\left(\frac{R }{t}\right)^2 
$. 
After some rearrangements, we finally get Eq.~\eqref{eqn: zero-pressure k, long cylinders}. \newpage 
\section{Simulation methods: shell indentation (\COMSOL{})}  \label{app: comsol} 
In this appendix, we provide implementation details of finite element simulations of the indentation studies (\textbf{\ref{subsec: the local indentation stiffness}}), which were performed using the software \texttt{COMSOL Multiphysics}. We used the \texttt{Stationary} solver with the \texttt{Shell} module to simulate the equilibrium configurations of orthotropic thin shells under combined pressure and point loads. Geometric nonlinearity was enabled to ensure that the influence of the pressure-induced prestress was correctly accounted for in the indentation study.  
\subsection{Orthotropic Materials} 
\texttt{COMSOL} allows for the definition of arbitrary anisotropic elastic materials using the \texttt{Material} module.
It is known that a three-dimensional orthotropic material has \emph{nine} independent elastic constants; these include 
three Young's moduli   ($E_1$,           $E_2$           and $E_3$), 
three Poisson's ratios ($\upsilon_{12}$, $\upsilon_{13}$ and $\upsilon_{23}$) and 
three shear   moduli   ($G_{12}$,        $G_{13}$        and $G_{23}$)~\cite{Anisotropy_Text}. 
The nine parameters have to satisfy constraints that stem from positive definiteness of the corresponding stiffness tensor. 
This makes it challenging to choose sets of these parameters which can guarantee stable simulations. 
We therefore followed the presentation by Li and Barbi\v{c} for simulating orthotropic materials~\cite{Li2014_Orthotropic_Materials}. 
The essence of their approach is summarized below. 
\par 
Li and Barbi\v{c} consider a subclass of orthotropic materials which can be characterized with only \emph{four} independent parameters: 
$
\left\{ 
E_1, 
\lambda \equiv \lambda_{12} \coloneqq \frac{E_1}{E_2}, 
               \lambda_{13} \coloneqq \frac{E_1}{E_3}, 
\upsilon_\eff 
\right\} 
$. 
The last parameter $\upsilon_\eff$ is related to the three Poisson's ratios in the following way: 
\begin{equation} 
          \upsilon_\eff 
\coloneqq \sqrt{ \upsilon_{12}\upsilon_{21} } 
   \Reqeq \sqrt{ \upsilon_{13}\upsilon_{31} } 
   \Reqeq \sqrt{ \upsilon_{23}\upsilon_{32} }, 
\end{equation} 
which implies (using the facts $\frac{ \upsilon_{ij} }{E_i} = \frac{ \upsilon_{ji} }{E_j}$) that 
\begin{equation} 
  \upsilon_{ij} 
= \upsilon_\eff\sqrt{ \frac{E_i}{E_j} } 
  \quad 
  ( i, j \in \{1, 2, 3\} ). 
\end{equation} 
The three shear moduli are given by the corresponding Huber form: 
\begin{equation} 
       G_{ij} 
\Reqeq \frac{ \sqrt{E_i E_j} }{ 2(1 + \upsilon_\eff) }. 
\end{equation} 
The positive definiteness constraints require that $E_1, \lambda, \lambda_{13} \in \mathbb R_{> 0}$, and $\upsilon_\eff \in ( -1, \frac{1}{2} ]$.\footnote{
Note that the isotropic Poisson's ratio $\upsilon_\text{iso}$ has the same range as $\upsilon_\eff$: $\upsilon_\text{iso} \in ( -1, \frac{1}{2} ]$. 
This is indeed the key motivation for introducing $\upsilon_\eff$. 
} 
\par 
In our simulations, we fixed the value of $E_1$ and $\upsilon_\eff$ to be $70\ \mathrm{GPa}$ and $0.3$, respectively. 
We also fixed the value of $\lambda_{13}$ after having verified that transverse shear deformations were indeed negligible in our studies. 
We chose $\lambda_{13} \Reqeq 2$. 
Therefore, in our simulations, there was really only \emph{one} free parameter that needed tuning to vary the degree of a thin shell's material anisotropy, namely $\lambda$. 
\subsection{Shells with Material Orthotropy and Boundary Conditions} 
The \texttt{3D Component} feature was first used to generate spheroidal and cylindrical surfaces. 
We then used the \texttt{Shell} module to turn these surfaces into actual shells. 
\subsubsection{Spheroidal Shells} \label{subsubsec: spheroids in COMSOL} 
A spheroid is an ellipsoid of revolution. 
To parametrize a spheroid, $\frac{x^2}{a^2} + \frac{y^2 + z^2}{b^2} = 1$, two parameters, $a$ and $b$, are needed. 
In our simulations, we fixed $b \Reqeq 1\ \mathrm{m}$ (so that $R_y \Reqeq 1\ \mathrm{m}$) and set $a \Reqeq \frac{b}{ \sqrt{1 - \beta_0} }$. 
We varied the asphericity of a spheroid by changing $\beta_0$ ($\beta_0 \in (-1, 1]$). 
The thickness of the spheroidal shell (denoted by $t$) was also fixed during each simulation. 
Since we were simulating thin shells, it is required that $\frac{b}{t} \gtrsim 50$. 
We used $t \Reqeq 1\ \mathrm{mm}$ in our simulations. 
\par 
For best results, we aimed for the mesh in the vicinity of the indentation point to be as fine as possible, relative to the characteristic length scales for thin-shell deflections which are the geometric means $\sqrt{R_1t}$ and $\sqrt{R_2t}$; however, setting the same fine mesh size for the entire shell was computationally impractical and also unnecessary: the main contribution of the rest of the shell away from the indentation region is to provide the geometry-determined  prestress in response to the internal pressure, which varies on much longer length scales of order $R_1$ and $R_2$.
Therefore, we assembled the shell surface out of separate regions with different mesh fineness requirements to balance physics performance with computational efficiency, as described below.

We used both the \texttt{Physics-controlled mesh} and the \texttt{User-controlled mesh} to build our spheroidal shells ($\beta_0 \neq 1$).
Each shell surface $S$ is composed of three disjoint regions: $S = S_\mathrm{top} \sqcup S_\mathrm{bot} \sqcup S_\mathrm{rest}$. 
Take the ellipsoid in \Figref{fig: orthotropic structures}\ (c) as an example. 
Among the three, the second region $S_\mathrm{bot}$ is centered at $O$. 
Its projection onto the tangent plane at $O$ is an elliptical disk $\mathcal E$ whose semi-major (semi-minor) axis is given by $\max\left\{3\sqrt{R_x t}, 3\sqrt{R_y t}\right\}$ ($\min\left\{3\sqrt{R_x t}, 3\sqrt{R_y t}\right\}$). 
The first region $S_\mathrm{top}$ is centered at the top of the ellipsoid but otherwise identical to $S_\mathrm{bot}$, and $S_\mathrm{rest}$ represents the rest of the shell surface. 
The first two regions can be built by obtaining the \texttt{Intersection} of $S$ and a \emph{solid} elliptical cylinder with cross-section $\mathcal E$ (the cylinder can be built with the built-in \texttt{Extrude} function), and the third region by taking the \texttt{Difference}. 
For $S_\mathrm{top}$ and $S_\mathrm{bot}$, we used the \texttt{User-controlled mesh} and set the mesh size \emph{exactly} to $3t = 3\ \mathrm{mm}$ (we enforced \texttt{Maximum element size} and \texttt{Minimum element size} to be equal). 
For $S_\mathrm{rest}$, the physics-controlled \texttt{Extremely fine} mesh size was used. 
\par 
Material orthotropy was implemented using the \texttt{Material} module. 
Orientations of material orthotropy were conveniently set, by default in \COMSOL, to coincide with the shell's \texttt{Global coordinate system}, which can be found under \texttt{Shell/Linear Elastic Material/Shell Local System/Coordinate System Selection/Coordinate system}. The default orientation recreated the desired alignment of the material directions with the symmetry directions of spheroidal shells for equatorial indentations (Fig. 1). For simulations on indentation response at the poles of an orthotropic sphere (\Appref{appendix: pole calculations}), we instead used the \texttt{Boundary System} for orienting the orthotropy directions, which conveniently put the two poles at the sphere's top and bottom. 
\par 
We used the boundary condition \texttt{Rigid Motion Suppression} for \texttt{All boundaries}. 
We also used the boundary condition \texttt{Symmetry} for \texttt{All edges} except the boundary of $S_\mathrm{top}$ and $S_\mathrm{bot}$. 
\par 
To simulate an internal pressure, a \emph{negative} \texttt{Face Load} was applied. 
For the zero-pressure simulations, the magnitude of the \texttt{Face Load} was set to zero.
The indentation itself was implemented using two instances of \texttt{Point Load} to ensure force balance: one \texttt{Point Load} with a \emph{negative} magnitude was applied at the top of the shell, $(0, 0, b)$, and a \emph{positive} one at the bottom, $(0, 0, -b)$. 
The absolute magnitude of the two loads was identical (this is essential for \texttt{Rigid Motion Suppression} to be used properly) and small, such that the resulting normal displacement, \texttt{shell.w}, was much less than the shell thickness $t$. 
In our simulations, the force magnitude used was $1\ \mathrm{N}$.
The ratio of the force magnitude to the resulting normal displacement provided the desired indentation stiffness measurement in our simulations.
\begin{figure}[H] 
\centering 
\includegraphics[width = 0.35\textwidth]{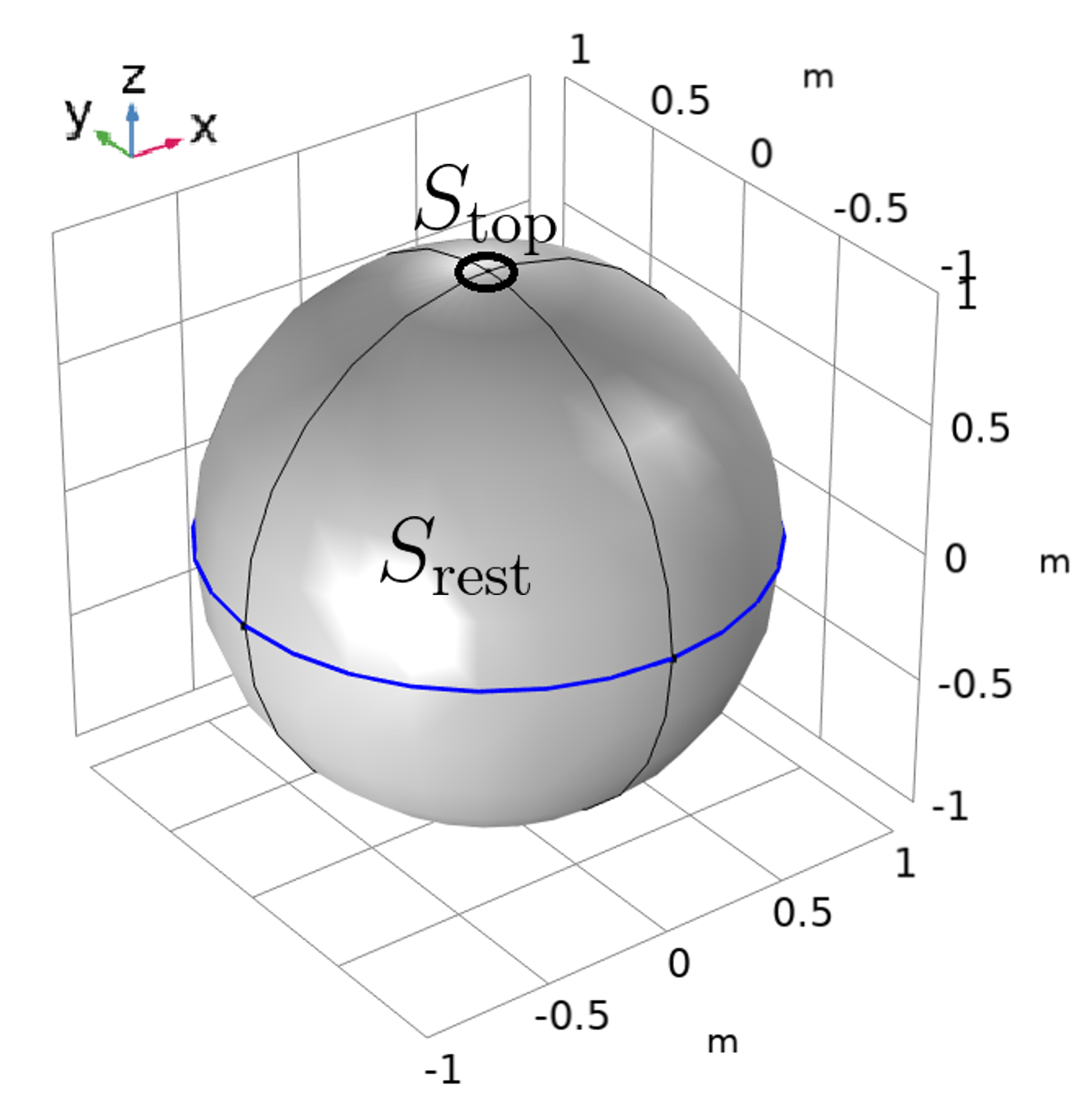} 
\caption{
\COMSOL{} setup for a sphere ($\beta_0 = 0$). The different mesh regions are indicated. 
} 
\label{fig appendix: COMSOL shell surface setup} 
\end{figure} 
\subsubsection{Cylindrical Shells} \label{subsubsec: cylinders in COMSOL} 
The radius of the cylindrical shells (denoted by $R$) was fixed to be $1\ \mathrm{m}$. 
As for spheroidal shells, it is required that $\frac{R}{t} \gtrsim 50$, where $t$ again denotes the shell thickness. 
As before, we used $1\ \mathrm{mm}$ for $t$. 
We also used the same loading conditions (\texttt{Face Load} and \texttt{Point Load}); however, depending on the magnitude of the internal pressure, different geometries with associated boundary conditions were employed. 
\paragraph{Low pressures.} Under this circumstance, which includes the zero-pressure case, indentation response of long cylinders is not localized. 
In our simulations, this corresponds to the pressure range $\p \lesssim 10^{-4}$. 
For this pressure range, we run our simulations with real cylindrical shells for which $\beta_0$ is exactly equal to one. 
Because \COMSOL{} is not able to simulate infinite cylinders, we set the length of our shells to be $10R\sqrt{ \frac{R}{t} }$. 
The combination $R\sqrt{ \frac{R}{t} }$ is the characteristic deformation length scale for indenting a cylinder at zero pressure~\cite{de_Pablo2003}: The indentation response becomes negligible at distances greater than this length scale away from the point load. 
For this geometry, \texttt{Rigid Motion Suppression} was again imposed for \texttt{All boundaries}, and \texttt{Symmetry} for \texttt{All edges}. 
We only used the \texttt{User-controlled mesh} to build our cylindrical shells; the mesh size belongs to the range $(10t, 1000t)$. 
\paragraph{High pressures.} By ``high pressure'' we mean that the internal pressure that a cylinder is subjected to is high enough, so that the resulting indentation response starts to become localized~\cite{Wenqian2021}, and it is accurate to use the double Fourier transform.
In our simulations, this happens when the scaled pressure is of the order of $10^{-2}$ ($\p \sim 10^{-2}$). However, we found that for perfectly cylindrical shells, the prestress components computed by \COMSOL{} did not match the well-established results for cylindrical pressure vessels~\cite{Timoshenko_Introductory}.
While we could not pinpoint the source of this discrepancy, we observed that the discrepancy was eliminated upon using highly elongated spheroids with $\beta_0 = 0.9999$, which have approximately the same prestress profile and the same local geometry as cylinders at the equator.
Therefore, for $\p \gtrsim 10^{-2}$, we run our simulations using elongated spheroids with $\beta_0 = 0.9999$. 
The \texttt{Rigid Motion Suppression} was again imposed for \texttt{All boundaries}. 
Unlike before, \texttt{Symmetry} was only imposed for the edges with respect to which the top and bottom of a spheroid are symmetric (e.g., the blue curves in \Figref{fig appendix: COMSOL shell surface setup}); imposing \texttt{Symmetry} for \texttt{All edges} created issues related to a known ``bursting'' instability of nearly cylindrical shells at very high pressures~\cite{Svensson2021_Bursting}.

We built the surface of these elongated spheroidal shells using the same three regions as before (see \Appref{subsubsec: spheroids in COMSOL}). 
For $S_\mathrm{top}$ and $S_\mathrm{bot}$, the \texttt{User-controlled mesh} was used with mesh size belonging to the range $(3t, 30t)$, and the physics-controlled \texttt{Extremely fine} mesh size was used for $S_\mathrm{rest}$. \newpage 
\section{Evaluating the Stiffness Integrals} \label{appendix: evaluating integrals} 
In this appendix, we will show the details how we evaluated the definite integrals in Eqs.~\eqref{eqn: pressurized k, spheroids} and~\eqref{eqn: pressurized k, long cylinders}. 
We will start with the latter, which is a special case of the former. 
\subsection{Equation~\eqref{eqn: pressurized k, long cylinders}} 
Setting $\beta_0 = 1$ (and hence $\beta' = 1$ and $\beta_\lambda' = 2\sqrt{\lambda} - 1$) in Eq.~\eqref{eqn: pressurized k, spheroids} gives 
\begin{equation} 
  \frac{1}{ k_\cyl(\p(\lambda), \lambda) } 
\coloneqq 
  \frac{1}{8\pi^2}\sqrt{ \frac{R'^2}{D' Y'} } 
  \mathcal{I}_1(\p(\lambda), \lambda), 
\end{equation} 
where $R' = \sqrt[4]{\lambda}R$ with $R$ the radius of cylinders, and 
\begin{align} 
\begin{split} 
    \mathcal{I}_1(\p(\lambda), \lambda) 
& \coloneqq 
    \int_0^{2\pi} \df{\varphi}\, 
    \int_0^{+\infty} 
    \frac{ \df{u} } 
    { 
    u^2 
  + 2\p\left(1 + \beta_\lambda'\sin^2{\varphi}\right)u 
  +                            \cos^4{\varphi} 
    } 
    \\[0.25 em] 
& = 4\int_0^\frac{\pi}{2} \df{\varphi}\, 
     \int_0^{+\infty} 
    \frac{ \df{u} } 
    { 
    u^2 
  + 2\p\left(1 + \beta_\lambda'\cos^2{\varphi}\right)u 
  +                            \sin^4{\varphi} 
    }; 
\end{split} 
\end{align} 
we have used the fact 
\begin{equation} 
   \int_0^{2\pi}        \df{\varphi}\, f( \cos^2{\varphi} ) 
= 4\int_0^\frac{\pi}{2} \df{\varphi}\, f( \sin^2{\varphi} ). 
\end{equation} 
To evaluate $\mathcal{I}_1(\p(\lambda), \lambda)$, we make two changes of variables: $s = u\csc^2{\varphi}$ and $t = \cot{\varphi}$; as a result, 
\begin{align} 
\begin{split} 
    \mathcal{I}_1(\p(\lambda), \lambda) 
& = 4\int_0^\frac{\pi}{2} \df{\varphi}\, \frac{ \csc^4{\varphi} }{ \csc^4{\varphi} }\, 
     \int_0^{+\infty} 
    \frac{ \df{u} } 
    { 
    u^2 
  + 2\p\left(1 + \beta_\lambda'\cos^2{\varphi}\right)u 
  +                            \sin^4{\varphi} 
    } 
    \\[0.25 em] 
& = 4\int_0^\frac{\pi}{2} \df{\varphi}\, \csc^2{\varphi}\, 
     \int_0^{+\infty} 
    \frac{ \df{ \left(u\csc^2{\varphi}\right) } } 
    { 
       \left(u\csc^2{\varphi}\right)^2 
  + 2\p\left( \csc^2{\varphi} + \beta_\lambda'\cot^2{\varphi}\right) 
       \left(u\csc^2{\varphi}\right) 
  + 1 
    } 
    \\[0.25 em] 
& = 4\int_0^{+\infty} \int_0^{+\infty} 
    \frac{ \df{s}\, \df{t} } 
    { 
    s^2 
  + 2\p\left(1 + 2\sqrt{\lambda}t^2\right)s 
  + 1 
    }, 
\end{split} 
\end{align} 
where in the last step, we changed the order of integration. 
We notice at this point that we can easily ``tease out'' the integral's explicit $\lambda$-dependence by making another change of variables 
$v = 2\sqrt{\p}\sqrt[4]{\lambda}t$: 
\begin{equation} 
  \mathcal{I}_1(\p(\lambda), \lambda) 
= \frac{2}{ \sqrt{\p}\sqrt[4]{\lambda} }\int_0^{+\infty} \int_0^{+\infty} 
  \frac{ \df{s}\, \df{v} }{ s v^2 + (s^2 + 2\p s + 1) }. 
\end{equation} 
It is now straightforward to evaluate $\mathcal{I}_1(\p(\lambda), \lambda)$: 
\begin{align} 
\begin{split} 
    \mathcal{I}_1(\p(\lambda), \lambda) 
& = \frac{2}{ \sqrt{\p}\sqrt[4]{\lambda} }\int_0^{+\infty} \frac{ \df{s} }{s}\, 
    \int_0^{+\infty} \frac{ \df{v} } 
    {\displaystyle 
    v^2 
  + \left(\sqrt{ \frac{s^2 + 2\p s + 1}{s} }\right)^2 
    } 
    \\[0.25 em] 
& = \frac{\pi}{ \sqrt{\p}\sqrt[4]{\lambda} }\int_0^{+\infty} \frac{ \df{s} }{ \sqrt{s} }\, 
    \frac{1}{ \sqrt{s^2 + 2\p s + 1} } 
    \\[0.25 em] 
& = \frac{2\pi}{ \sqrt{\p}\sqrt[4]{\lambda} }\int_0^{+\infty} 
    \frac{ \df{x} }{ \sqrt{x^4 + 2\p x^2 + 1} }. 
\end{split} 
\end{align} 
The last integral in the equation above can be expressed in terms of the complete elliptic integral of the first kind~\cite{Ryzhik_Integral_Table}: 
\begin{align} 
\begin{split} 
    \int_0^{+\infty} 
    \frac{ \df{x} }{ \sqrt{x^4 + 2\p x^2 + 1} } 
& = \int_0^{+\infty} 
    \frac{ \df{x} } 
    { 
    \sqrt{ (x^2 + 1)^2 - 2(1 - \p)x^2 } 
    } 
    \\[0.25 em] 
& = \int_0^\frac{\pi}{2} 
    \frac{ \df{ ( \tan{\theta} ) } } 
    { 
    \sqrt{ (\tan^2{\theta} + 1)^2 - 2(1 - \p)\tan^2{\theta} } 
    } 
  = \int_0^\frac{\pi}{2} 
    \frac{ \sec^2{\theta}\, \df{\theta} } 
    { 
    \sqrt{ \sec^4{\theta} - 2(1 - \p)\tan^2{\theta} } 
    } 
    \\[0.25 em] 
& = \int_0^\frac{\pi}{2} 
    \frac{ \df{\theta} } 
    { 
    \sqrt{ 
    \displaystyle 
    1 - \frac{1}{2}(1 - \p)\sin^2{2\theta} 
    } 
    } 
  = \frac{1}{2}\int_0^\pi 
    \frac{ \df{\phi} } 
    { 
    \sqrt{ 
    \displaystyle 
    1 - \frac{1}{2}(1 - \p)\sin^2{\phi} 
    } 
    } 
    \\[0.25 em] 
& = \EllipticK{ \frac{1}{2}(1 - \p) }; 
\end{split} 
\end{align} 
to arrive at the last step, we have used 
\begin{equation} 
   \int_0^{\pi}         \df{\phi}\, g( \sin^2{\phi} ) 
= 2\int_0^\frac{\pi}{2} \df{\phi}\, g( \sin^2{\phi} ). 
\end{equation} 
After some more algebra, we obtain Eq.~\eqref{eqn: pressurized k, long cylinders}. 
\par\bigskip 
\begin{remark} 
By changing the order of integration, like what we did in Ref.~\onlinecite{Wenqian2021}, we get the following identity for the complete elliptic integral of the first kind: 
\begin{equation} 
\boxed{ 
  \EllipticK{x} 
= \frac{ \sqrt{2} }{\pi}\int_0^{+\infty} \df{u}\, 
  \frac 
  { \arccos   \left(u^2 + 1 - 2x\right)    } 
  { \sqrt{1 - \left(u^2 + 1 - 2x\right)^2} } 
}. 
\end{equation} 
\end{remark} 
\subsection{Equation~\eqref{eqn: pressurized k, spheroids}} 
We now return to the more general case, Eq.~\eqref{eqn: pressurized k, spheroids}; the way of evaluating the integral is essentially the same, but the changes of variables involved will require slightly more thoughts. 
\par 
We start by rewriting the integral in Eq.~\eqref{eqn: pressurized k, spheroids}: 
\begin{align} 
\begin{split} 
    \mathcal{I}(\p(\lambda), \beta_0, \lambda) 
& \coloneqq 
    \int_0^{2\pi} \df{\varphi}\, 
    \int_0^{+\infty} 
    \frac{ \df{u} } 
    { 
    u^2 
  + 2\p\left(1 + \beta_\lambda'\sin^2{\varphi}\right)  u 
  +    \left(1 - \beta        '\sin^2{\varphi}\right)^2 
    } 
    \\[0.25 em] 
& = 4\int_0^\frac{\pi}{2} \df{\varphi}\, 
     \int_0^{+\infty} 
    \frac{ \df{u} } 
    { 
    u^2 
  + 2\p\left(1 + \beta_\lambda'\cos^2{\varphi}\right)  u 
  +    \left(1 - \beta        '\cos^2{\varphi}\right)^2 
    } 
    \\[0.25 em] 
& = 4\int_0^\frac{\pi}{2} \frac{ \df{\varphi} }{ 1 - \beta'\cos^2{\varphi} }\, 
     \int_0^{+\infty} 
    \frac{ \df{v} } 
    { 
    \displaystyle 
    v^2 
  + 2\p\left( 
    \frac 
    { 1 + \beta_\lambda'\cos^2{\varphi} } 
    { 1 - \beta        '\cos^2{\varphi} } 
    \right) 
    v 
  + 1 
    }, 
\end{split} 
\end{align} 
where $v \coloneqq \frac{u}{ 1 - \beta'\cos^2{\varphi} }$. 
Realizing 
\begin{equation} 
  \odfrac{}{\varphi}{ \arctan\left( 
  \frac{1}{ \sqrt{1 - \beta'} }\tan{\varphi} 
  \right) } 
= \sqrt{    1 - \beta'} 
  \frac{1}{ 1 - \beta'\cos^2{\varphi} }, 
\end{equation} 
we make the change of variables 
\begin{equation} 
s = \arctan\left( 
    \frac{1}{ \sqrt{1 - \beta'} }\tan{\varphi} 
    \right). 
\end{equation} 
As a result, we can make the following simplification: 
\begin{align} 
\begin{split} 
    \frac 
    { 1 + \beta_\lambda'\cos^2{\varphi} } 
    { 1 - \beta        '\cos^2{\varphi} } 
& = 1 
  + \frac 
    {\beta' + \beta_\lambda'} 
    {\sec^2{\varphi} - \beta'} 
    \\[0.25 em] 
& = 1 
  + \frac 
    {\beta' + \beta_\lambda'} 
    {1 + (1 - \beta')\tan^2{s} - \beta'} 
  = 1 
  + \frac 
    {\beta' + \beta_\lambda'} 
    {1 - \beta'} 
    \cos^2{s} 
  \coloneqq 
    1 + \alpha'\cos^2{s}, 
\end{split} 
\end{align} 
where 
\begin{equation} 
  \alpha' 
\coloneqq 
  \frac 
  {\beta' + \beta_\lambda'} 
  {1 - \beta'} 
= \frac 
  {   2\beta_0} 
  {1 - \beta_0} 
= \frac 
  {1 + \beta_0} 
  {1 - \beta_0} 
-  1 
\coloneqq 
  \alpha - 1, 
\end{equation} 
a combination of parameters, which is independent of $\lambda$; accordingly, with the new integration variable, 
\begin{equation} 
  \mathcal{I}(\p(\lambda), \beta_0, \lambda) 
= \frac{4}{ \sqrt{1 - \beta'} } 
  \int_0^\frac{\pi}{2} \df{s}\, 
  \int_0^{+\infty} 
  \frac{ \df{v} } 
  { 
  v^2 
+ 2\p\left( 
  1 + \alpha'\cos^2{s} 
  \right) 
  v 
+ 1 
  }. 
\end{equation} 
We notice that all the \emph{explicit} $\lambda$-dependence is in the prefactor $\frac{4}{ \sqrt{1 - \beta'} }$. 
\par 
Changing the order of integration, we first evaluate the $s$-integral. 
After some algebra, we arrive at 
\begin{equation} 
  \mathcal{I}(\p(\lambda), \beta_0, \lambda) 
= \frac{2\pi}{ \sqrt{1 - \beta'} } 
  \int_0^{+\infty} 
  \frac{ \df{v} }{v^2 + 2\p v + 1} 
  \frac{1} 
  { 
  \displaystyle 
  \sqrt{ 
  1 
+ \frac{2\alpha'\p v}{v^2 + 2\p v + 1} 
  } 
  }, 
\end{equation} 
where we have used 
\begin{equation} 
  \int_0^\frac{\pi}{2} \frac{ \df{s} }{ A + B\cos^2{s} } 
= \frac{\pi}{2}\frac{1}{ \sqrt{A} }\frac{1}{ \sqrt{A + B} }. 
\end{equation} 
We notice that the term 
$
  \left( 
  1 
+ \frac{2\alpha'\p v}{v^2 + 2\p v + 1} 
  \right)^{ -\frac{1}{2} } 
$
contains all the non-trivial geometric dependence: Setting $\beta_0 = 0$ ($\beta' = 1 - \sqrt{\lambda}$ and $\alpha' = 0$) gives 
\begin{equation} 
  \mathcal{I}(\p(\lambda), \beta_0 = 0, \lambda) 
= \frac{2\pi}{ \sqrt[4]{\lambda} } 
  \int_0^{+\infty} 
  \frac{ \df{v} }{v^2 + 2\p v + 1} 
\end{equation} 
which is the familiar integral corresponding to the stiffness of a spherical shell. 
\par 
Realizing 
\begin{equation} 
  \odfrac{}{v}{ 
  \arctan\left( 
  \frac{v + \p}{ \sqrt{1 - \p^2} } 
  \right) 
  } 
= \sqrt{1 - \p^2} 
  \frac{1}{v^2 + 2\p v + 1}, 
\end{equation} 
we make the change of variables 
\begin{equation} 
t = \arctan\left( 
    \frac{v + \p}{ \sqrt{1 - \p^2} } 
    \right). 
\end{equation} 
As a consequence, we can rewrite the term just mentioned, which is related to geometric anisotropy, in terms of $t$: 
\begin{align} 
\begin{split} 
    \frac{v}{v^2 + 2\p v + 1} 
& = \frac 
    { \sqrt{1 - \p^2}\tan{t} - \p} 
    { \sqrt{1 - \p^2} } 
    \left(\odfrac{v}{t}(t)\right)^{-1} 
  = \frac 
    {\sqrt{1 - \p^2}\tan  {t} - \p} 
    {     (1 - \p^2)\sec^2{t} } 
    \\[0.25 em] 
& = -\frac{1}{ 2(1 - \p^2) }\left( 
  - \sqrt{1 - \p^2}\sin{2t} 
  +           \p   \cos{2t} 
  +           \p 
    \right) 
    \\[0.25 em] 
& = -\frac{1}{ 2(1 - \p^2) }\left[ 
    \cos\left(2t + \arccos{\p}\right) 
  + \p 
    \right]. 
\end{split} 
\end{align} 
It follows that in terms of the new integration variable, 
\begin{align} 
\begin{split} 
    \mathcal{I}(\p(\lambda), \beta_0, \lambda) 
& = \frac{2\pi}{ \sqrt{1 - \beta'} }\frac{1}{ \sqrt{1 - \p^2} } 
    \int_{ \arcsin{\p} }^\frac{\pi}{2} 
    \frac{ \df{t} } 
    {\displaystyle 
    \sqrt{ 
    1 
  - \alpha'\frac{\p}{1 - \p^2}\left[ 
    \cos\left(2t + \arccos{\p}\right) 
  + \p 
    \right] 
    } 
    } 
    \\[0.25 em] 
& = \frac{\pi}{ \sqrt{1 - \beta'} }\frac{1}{ \sqrt{1 - \p^2} } 
    \int_{ \pi - \arccos{\p} }^{ \pi + \arccos{\p} } 
    \frac{ \df{\theta} } 
    {\displaystyle 
    \sqrt{ 
    1 
  - \alpha'\frac{\p^2}{1 - \p^2} 
  - \alpha'\frac{\p  }{1 - \p^2} 
    \cos{\theta} 
    } 
    }, 
\end{split} 
\end{align} 
where we used the following identities: 
\begin{equation} 
  \arctan\left( 
  \frac{\p}{ \sqrt{1 - \p^2} } 
  \right) 
= \arcsin{\p} 
  \textAnd 
  \arccos{\p} 
+ \arcsin{\p} 
= \frac{\pi}{2}, 
\end{equation} 
and we also changed the integration variable from $t$ to $\theta = 2t + \arccos{\p}$. 
Performing another change of variables $\phi = \theta - \pi$, we can rewrite $\mathcal{I}(\p(\lambda), \beta_0, \lambda)$ as follows: Factoring out the term 
$
  \sqrt{ 
  1 
- \alpha'\frac{\p^2}{1 - \p^2} 
  } 
$
from the denominator of the integrand, 
\begin{equation} 
  \mathcal{I}(\p(\lambda), \beta_0, \lambda) 
= \frac{2\pi}{ \sqrt{1 - \beta'} }\frac{1}{ \sqrt{1 - \alpha\p^2} } 
  \int_0^{ \arccos{\p} } 
  \frac{ \df{\phi} } 
  {\displaystyle 
  \sqrt{ 
  1 
+ \frac{\alpha'\p}{1 - \alpha\p^2} 
  \cos{\phi} 
  } 
  }; 
\end{equation} 
recall $\alpha = \alpha' + 1$. 
Using the identity 
\begin{equation} 
  \int_0^\vartheta 
  \frac{ \df{\phi} }{ \sqrt{ 1 + A\cos{\phi} } } 
= \frac{2}{ \sqrt{1 + A} } 
  \EllipticF{\frac{1}{2}\vartheta}{ \frac{2A}{1 + A} }, 
\end{equation} 
we finally get, after some rewriting, Eq.~\eqref{eqn: pressurized k, spheroids, the closed form}. \newpage 
\section{Simulation methods: buckling of orthotropic spheres} \label{app:bucklingsim}

The buckling pressure of orthotropic spherical shells, \Figref{fig: Buckling pressure of orthotropic spheres}, was determined with finite element simulations. As the implementation in C++ is based on previous work \cite{Herrmann2013, Herrmann2015, Vetter2015, Herrmann2019}, we summarize only the main aspects here.

Denote by $\Omega \subset \mathbb{R}^3$ the middle surface of the thin shell with thickness $t$. We now distinguish between the stress-free reference configuration denoted by barred symbols, and the deformed configuration, denoted by bare symbols. Thus, $\overline{\Omega} \subset \mathbb{R}^3$ is the middle surface of the unstrained shell (a sphere in the case considered here). We describe the shell in a total Lagrangian formulation, with $\overline{{\bf x}}(x^1,x^2)$ and ${\bf x}(x^1,x^2)$ curvilinear parameterizations of $\overline{\Omega}$ and $\Omega$, respectively. The tangent spaces of $\overline{\Omega}$ and $\Omega$ are then spanned by
\begin{equation}
\overline{{\bf a}}_{\alpha}(x^1,x^2) = \overline{{\bf x}}_{,\alpha} = \frac{\partial\overline{\bf x}}{\partial x^{\alpha}},\qquad {\bf a}_{\alpha}(x^1,x^2) = {\bf x}_{,\alpha} = \frac{\partial\bf x}{\partial x^{\alpha}},
\end{equation}
and by virtue of the Kirchhoff assumption, the shell directors are given by the unit surface normals
\begin{equation}
{\overline{\bf a}}_3 = \frac{{\overline{\bf a}}_1\times {\overline{\bf a}}_2}{\norm{{\overline{\bf a}}_1\times {\overline{\bf a}}_2}},\qquad {\bf a}_3 = \frac{{\bf a}_1\times {\bf a}_2}{\norm{{\bf a}_1\times {\bf a}_2}}.
\end{equation}
To define the membrane and bending strains, we require the covariant components of the metric tensor,
\begin{equation}
\overline{a}_{\alpha\beta} = \overline{{\bf a}}_{\alpha} \cdot \overline{{\bf a}}_{\beta},\qquad a_{\alpha\beta} = {\bf a}_{\alpha} \cdot {\bf a}_{\beta},
\end{equation}
and those of the shape tensor,
\begin{equation}
\overline{b}_{\alpha\beta} = \overline{{\bf a}}_{3} \cdot \overline{{\bf a}}_{\alpha,\beta},\qquad b_{\alpha\beta} = {\bf a}_{3} \cdot {\bf a}_{\alpha,\beta}.
\end{equation}
Since the thin shell is in a state of locally plane stress, the strain tensors for stretching and bending with respect to the curvilinear coordinates can be expressed in Voigt notation as
\begin{equation}
\boldsymbol{\mathbf \alpha} = \begin{pmatrix}\alpha_{11}\\\alpha_{22}\\2\alpha_{12}\end{pmatrix} = \frac{1}{2}\begin{pmatrix}a_{11}-\overline{a}_{11}\\a_{22}-\overline{a}_{22}\\2(a_{12}-\overline{a}_{12})\end{pmatrix},\qquad
\boldsymbol{\mathbf \beta} = \begin{pmatrix}\beta_{11}\\\beta_{22}\\2\beta_{12}\end{pmatrix} = \begin{pmatrix}\overline{b}_{11}-b_{11}\\\overline{b}_{22}-b_{22}\\2(\overline{b}_{12}-b_{12})\end{pmatrix}.
\end{equation}
We now transform these into an orthonormal basis $\{\boldsymbol{\mathbf e}_1,\boldsymbol{\mathbf e}_2\}$ of the tangent space, with respect to which the material orthotropy is expressed, using a transformation matrix $\mathbf{T}$ \cite{Vetter2015}:
\begin{equation}
\boldsymbol{\mathbf \varepsilon} = \mathbf{T} \boldsymbol{\mathbf \alpha},\qquad
\boldsymbol{\mathbf \kappa} = \mathbf{T} \boldsymbol{\mathbf \beta}
\end{equation}
with
\begin{equation}
\mathbf{T} = \begin{pmatrix}
t_{11}^2 & t_{21}^2 & t_{11}t_{21}\\
t_{12}^2 & t_{22}^2 & t_{12}t_{22}\\
2t_{11}t_{12} & 2t_{21}t_{22} & t_{11}t_{22}+t_{12}t_{21}
\end{pmatrix},\qquad t_{\alpha\beta} = \overline{\boldsymbol{\mathbf a}}^\alpha\cdot\boldsymbol{\mathbf e}_\beta.
\end{equation}
For a spherical shell, we define the material coordinate system aligned with the polar and azimuthal directions:
\begin{equation}
\boldsymbol{\mathbf e}_1 = \frac{\overline{\boldsymbol{\mathbf x}}}{\norm{\overline{\boldsymbol{\mathbf x}}}}\times\boldsymbol{\mathbf e}_2,\qquad\boldsymbol{\mathbf e}_2 = \frac{\unitvec{z}\times\overline{\boldsymbol{\mathbf x}}}{\norm{\unitvec{z}\times\overline{\boldsymbol{\mathbf x}}}}
\end{equation}
where $\unitvec{z}=\VEC{0,0,1}^\top$.

With these definitions, the potential energy of a pressurized, orthotropic thin shell can be expressed as \cite{Vetter2015}
\begin{equation}
U = \int_{\overline{\Omega}}\frac{1}{2}\left( t\boldsymbol{\mathbf \varepsilon}^\top\mathbf{C}\boldsymbol{\mathbf \varepsilon}+\frac{t^3}{12}\boldsymbol{\mathbf \kappa}^\top\mathbf{C}\boldsymbol{\mathbf \kappa}\right) - p\,\boldsymbol{\mathbf a}_3\cdot(\boldsymbol{\mathbf x}-\overline{\boldsymbol{\mathbf x}}) \,d\overline{\Omega}
\end{equation}
where $p$ is the internal-to-external pressure difference, $d\overline{\Omega} = \norm{\overline{{\bf a}}_1 \times \overline{{\bf a}}_2}\;dx^1 dx^2$ the reference area element, and
\begin{equation}
\mathbf{C} = \begin{pmatrix}E_1/(1-\upsilon_{12}\upsilon_{21}) & \upsilon_{21}E_1/(1-\upsilon_{12}\upsilon_{21}) & 0\\\upsilon_{12}E_2/(1-\upsilon_{12}\upsilon_{21}) & E_2/(1-\upsilon_{12}\upsilon_{21}) & 0\\0 & 0 & G_{12}\end{pmatrix}
\end{equation}
the elastic tensor for orthotropic plane stress.
To minimize $U$ numerically, we discretized the spherical shell into a triangulated mesh that was built by recursively subdividing a regular icosahedron five times, resulting in a so-called ``icosphere'' consisting of 20480 triangles and 10242 vertices. 10\% of the average edge length was added to each vertex position on the sphere as random noise to break the mesh symmetry. Using $C^1$-conforming Loop subdivision surface shape functions \cite{Cirak2000}, the middle surface can then be expressed as linear combinations of the shape functions $N_I$ with the nodal positions $x_I$ as weights:
\begin{equation}
\overline{\boldsymbol{\mathbf x}}(x^1,x^2) = \sum_{I=1}^{12}\overline{\boldsymbol{\mathbf x}}_I N_I,\qquad \boldsymbol{\mathbf x}(x^1,x^2) = \sum_{I=1}^{12}\boldsymbol{\mathbf x}_I N_I.
\end{equation}
(Note that for evaluation of the surface on patches with nodes of valence other than six, a recursive procedure is needed \cite{Cirak2000}.)
With this finite element discretization, and using a single Gauss point per triangle, the nodal forces can be assembled as \cite{Vetter2015}
\begin{equation}
\boldsymbol{\mathbf f}_I = -\sum_{e}\left\{\left(t\mathbf{M}_I^\top\mathbf{\hat{C}}\boldsymbol{\mathbf \alpha}+\frac{t^3}{12}\mathbf{B}_I^\top\mathbf{\hat{C}}\boldsymbol{\mathbf \beta}-pN_I\,\boldsymbol{\mathbf a}_3\right)\frac{\norm{\overline{{\bf a}}_1 \times \overline{{\bf a}}_2}}{2}\right\}_e,
\end{equation}
where the sum runs over all triangles $e$ within the local support of $N_I$, $\left\{\cdot\right\}_e$ denotes evaluation at the barycenter of $e$,
\begin{equation}
\mathbf{\hat{C}}=\mathbf{T}^\top\mathbf{C}\mathbf{T}
\end{equation}
is the elastic tensor transformed to the local frame, and $\mathbf{M}_I$ and $\mathbf{B}_I$ are membrane and bending matrices, whose transpose are column-wise given by \cite{Vetter2015}
\begin{align}
\mathbf{M}_I^\top &=
\begin{pmatrix}
N_{I,1}\boldsymbol{\mathbf a}_1 & N_{I,2}\boldsymbol{\mathbf a}_2 & N_{I,1}\boldsymbol{\mathbf a}_2 + N_{I,2}\boldsymbol{\mathbf a}_1
\end{pmatrix}\\
\mathbf{B}_I^\top &=
\begin{pmatrix}
\boldsymbol{\mathbf b}^I_{11} & \boldsymbol{\mathbf b}^I_{22} & 2\boldsymbol{\mathbf b}^I_{12}
\end{pmatrix}
\end{align}
with
\begin{equation}
\boldsymbol{\mathbf b}^I_{\alpha\beta} = \frac{1}{\norm{\boldsymbol{\mathbf a}_1\times\boldsymbol{\mathbf a}_2}}\left(\boldsymbol{\mathbf a}_{\alpha,\beta} - b_{\alpha\beta}\,\boldsymbol{\mathbf a}_3\right)\times\left(N_{I,1}\boldsymbol{\mathbf a}_2-N_{I,2}\boldsymbol{\mathbf a}_1\right) - N_{I,\alpha\beta}\boldsymbol{\mathbf a}_3.
\end{equation}

With the nodal forces, we solved Newton's equations of motion with far-subcritical viscous damping added, using a Newmark predictor-corrector method \cite{Vetter2015}. To determine the critical pressure, we slowly ramped up the applied pressure $p$ in the simulations until the shell collapsed.

\end{document}